\documentclass[superscriptaddress,preprintnumbers,byrevtex,floatfix]{revtex4}
\usepackage{graphics}
\usepackage{amssymb}
\usepackage{array}
\usepackage{verbatim}
\usepackage[ansinew]{inputenc}
\usepackage[english]{babel}
\usepackage{bm}
\usepackage{natbib}
\usepackage{hyperref}
\usepackage{amsmath}
\usepackage{subfigure}
\DeclareGraphicsExtensions{.eps,.tiff} 
 
\begin{document}

\title{Microfabricated Ion Traps}
\author{Marcus D. Hughes\footnote{Corresponding author. Email: M.D.Hughes@sussex.ac.uk}}
\author{Bjoern Lekitsch}
\author{Jiddu A. Broersma}
\author{Winfried K. Hensinger}
\affiliation{Department of Physics and Astronomy, University of Sussex, Brighton, UK\\
BN1 9QH}

\begin{abstract}

Ion traps offer the opportunity to study fundamental quantum systems with high level of accuracy highly decoupled from the environment. Individual atomic ions can be controlled and manipulated with electric fields, cooled to the ground state of motion with laser cooling and coherently manipulated using optical and microwave radiation. Microfabricated ion traps hold the advantage of allowing for smaller trap dimensions and better scalability towards large ion trap arrays also making them a vital ingredient for next generation quantum technologies. Here we provide an introduction into the principles and operation of microfabricated ion traps. We show an overview of material and electrical considerations which are vital for the design of such trap structures. We provide guidance in how to choose the appropriate fabrication design, consider different methods for the fabrication of microfabricated ion traps and discuss previously realized structures. We also discuss the phenomenon of anomalous heating of ions within ion traps, which becomes an important factor in the miniaturization of ion traps.\\

\textbf{Keywords:} Ion traps, microfabrication, quantum information processing, anomalous heating, laser cooling and trapping.
\bigskip

\end{abstract}

\maketitle

\section{Introduction}
Ion trapping was developed by Wolfgang Paul \cite{Paul} and Hans Dehmelt \cite{Dehmelt} in the 1950's and 60's and ion traps became an important tool to study important physical systems such as ion cavity QED \cite{Keller,Drewsen}, quantum simulators \cite{Pons,Friedenauer,Johanning,Clark,Kim}, determine frequency standards \cite{Udem,Webster,Tamm2,Chwalla}, as well as the development towards a quantum information processor \cite{Ciracgate,Wineland,Haeffner}. In general, ion traps compare well to other physical systems with good isolation from the environment and long coherence times. Progress in many of the research areas where ion traps are being used may be aided by the availability of a new generation of ion traps with the ion - electrode distance on the order of tens of micrometers. While in some cases, the availability of micrometer scale ion - electrode distance and a particular electrode shape may be of sole importance, often the availability of versatile and scalable fabrication methods (such as micro-electromechanical systems (MEMS) and other microfabrication technologies) may be required in a particular field. \\

One example of a field which will see step-changing innovation due to the emergence of microfabricated ion traps is the general area of quantum technology with trapped ions. In 1995 David DiVincenzo set out criteria which determine how well a system can be used for quantum computing \cite{DiVincenzo}. Most of these criteria have been demonstrated with an ion trap: Qubit initialization \cite{Leibfriedstate,Leibfriedstate2}, a set of universal quantum gates creating entanglement between ions \cite{Leibfriedphasequbit,Ciracgate,SorensenMolmer,Benhelm}, long coherence times \cite{Lucas}, detection of states \cite{Myersonreadout,acton} and a scalable architecture to host a large number of qubits \cite{Stick,Seidelin,Hensinger,Blakestad}. An important research area is the development of a scalable architecture which can incorporate all of the DiVincenzo criteria. A realistic architecture has been proposed consisting of an ion trap array incorporating storage and gate regions \cite{CiracandZoller,Kielpinski,Steane} and could be implemented using microfabricated ion traps. Microfabricated ion traps hold the possibility of small trap dimensions on the order of tens of micrometers and more importantly fabrication methods like photolithography that allow the fabrication of very large scale arrays. Electrodes with precise shape, size and geometry can be created through a number of process steps when fabricating the trap. \\

In this article we will focus on the design and fabrication of microfabricated radio frequency (rf) ion traps as a promising tool for many applications in ion trapping. Another ion trap type is the Penning trap \cite{Thompson} and advances in their fabrication have been discussed by Castrej\'{o}n-Pita et al. \cite{Castre} and will not be discussed in this article. Radio-frequency ion traps include multi-layer designs where the ion is trapped between electrodes located in two or more planes with the electrodes symmetrically surrounding the ion, for example as the ion trap reported by Stick et al. \cite{Stick}. We will refer to such traps as symmetric ion traps. Geometries where all the electrodes lie in a single plane and the ion is trapped above that plane, for example as the trap fabricated by Seidelin et al. \cite{Seidelin}, will be referred to as asymmetric or surface traps.\\

There have been many articles discussing ion traps and related physics, including studies of fundamental physics \cite{Paul2,Ghosh,Horvath}, spectroscopy \cite{Thompson2}, and coherent control \cite{Wineland}. This article focuses on the current progress and techniques used for the realisation of microfabricated ion traps. First we discuss basic principles of ion traps and their operation in section \ref{iontraps}. Section \ref{linear} discusses linear ion trap geometries as a foundation of most ion trap arrays. The methodology of efficient simulation of electric fields within ion trap arrays is discussed in Section \ref{simu}. Section \ref{ElecChara} discusses some material characteristics that have to be considered when designing microfabricated ion traps including electric breakdown and rf dissipation. Section \ref{FabProcess} provides a guide to realizing such structures with the different processes outlined together with the capabilities and limitations of each one. Finally in Section \ref{heating} we discuss motional heating of the ion due to fluctuating voltage patches on surfaces and its implications for the design and fabrication of microfabricated ion traps.

\section{Radio frequency ion traps}\label{iontraps}
Static electric fields alone cannot confine charged particles, this is a consequence of Earnshaw's theorem \cite{G99} which is analogous to Maxwell's equation $\nabla \cdot E=0$. To overcome this Penning traps use a combination of static electric and magnetic fields to confine the ion \cite{Penning,Thompson}. Radio frequency (rf) Paul traps use a combination of static and oscillating electric fields to achieve confinement. We begin with an introduction into the operation of radio frequency ion traps, highlighting important factors when considering the design of microfabricated ion traps.

\subsection{Ion trap dynamics}
\begin{figure}[htp]
\begin{center}
\resizebox*{8cm}{!}{\includegraphics{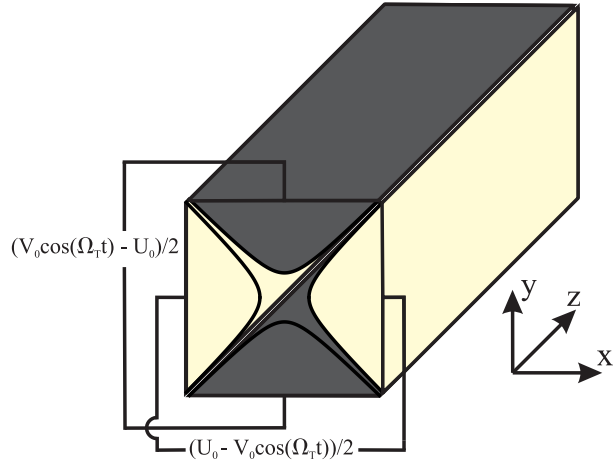}}%
\caption{Hyperbolic electrodes within the $x$ and $y$-axes where an rf voltage of $V_0\cos{(\Omega_Tt)}$ together with a static voltage of $U_0$ is applied to two opposite electrodes. The polarity of this voltage is reversed and applied to the other set of electrodes.}  
\label{linhyp}
\end{center}
\end{figure}

\begin{figure}[htp]
\begin{center}
\subfigure[]{
\resizebox*{5cm}{!}{\includegraphics{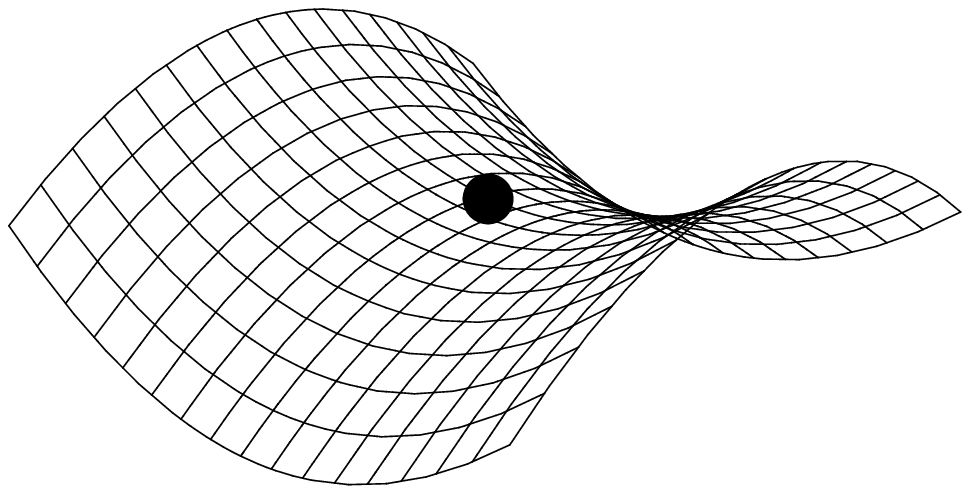}}}%
\subfigure[]{
\resizebox*{5cm}{!}{\includegraphics{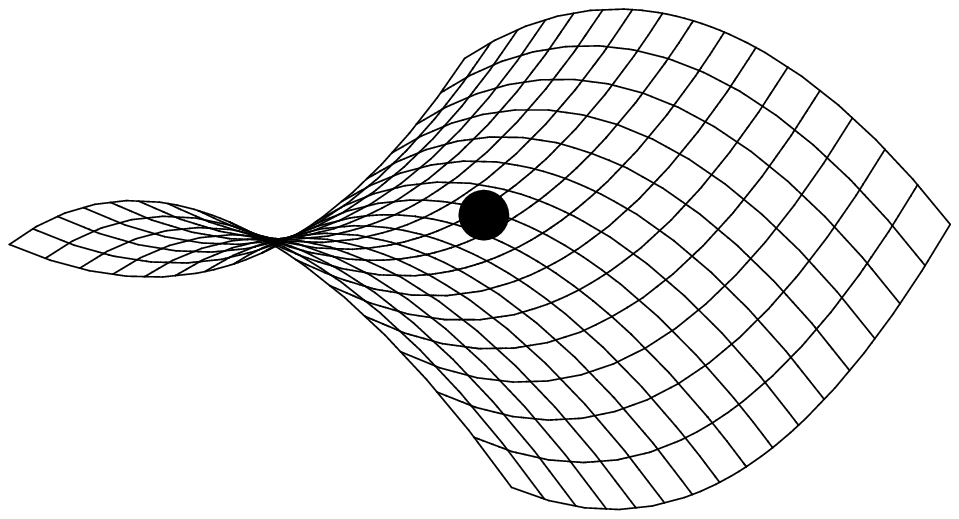}}}%
\subfigure[]{
\resizebox*{5cm}{!}{\includegraphics{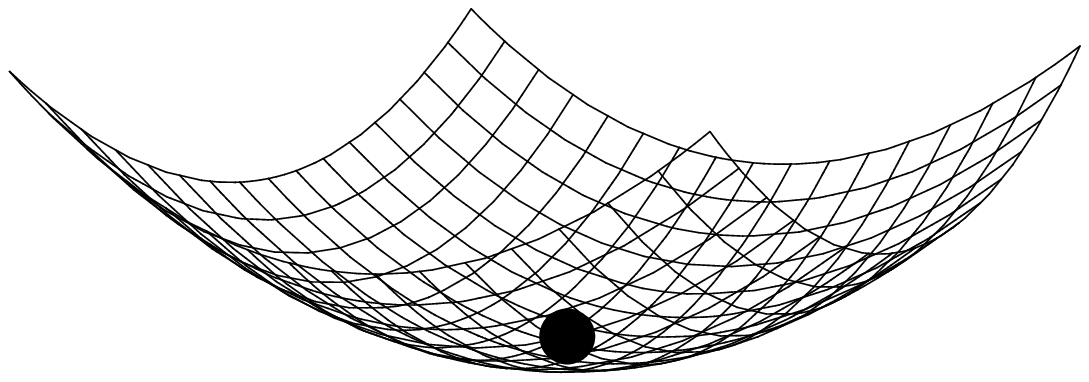}}}%
\caption{Principles of confinement with a pseudopotential. (a) A saddle potential created by a static electric field from a hyperbolic electrode geometry. (b) The saddle potential acquiring an inversion from a change in polarity. (c) The effective potential the ion sees resulting from the oscillating electric potential.}
\label{oscillating potential}
\end{center}
\end{figure}

First we consider a quadrupole potential within the radial directions, $x$ and $y$-axes, which is created from hyperbolic electrodes as shown in Fig. \ref{linhyp}, whilst there is no confinement within the axial ($z$) direction. By considering a static voltage $V_0$ applied to two opposite electrodes, the resultant electric potential will produce a saddle as depicted in Fig. \ref{oscillating potential} (a). With the ion present within this potential, the ion will feel an inward force in one direction and outward force in the direction perpendicular to the first. Reversing the polarity of the applied voltage the saddle potential will undergo an inversion as shown in Fig. \ref{oscillating potential} (b). As the force acting on the ion is proportional to the gradient of the potential, the magnitude of the force is less when the ion is closer to the centre. The initial inward force will move the ion towards the centre where the resultant outward force half a cycle later will be smaller. Over one oscillation the ion experiences a greater force towards the centre of the trap than outwards resulting in confinement. The effective potential the ion sees when in an oscillating electric field is shown in Fig. \ref{oscillating potential} (c). If the frequency of the oscillating voltage is too small then the ion will not be confined long enough in one direction. For the case where a high frequency is chosen then the effective difference between the inward and outward forces decreases and the resultant potential is minimal. By selecting the appropriate frequency $\Omega_T$ of this oscillating voltage together with the amplitude $V_0$ which is dependent on the mass of the charged particle, confinement of the particle can be achieved within the radial directions.\\
There are two ways to calculate the dynamics of an ion within a Paul trap, firstly a comprehensive treatment can be given using the Mathieu equation. The Mathieu equation provides a complete solution for the dynamics of the ion. It also allows for the determination of parameter regions of stability where the ion can be trapped. These regions of stability are determined by trap parameters such as voltage amplitude and rf drive frequency. Here we outline the process involved in solving the equation of motion via the Mathieu equation approach. By applying an oscillating potential together with a static potential, the total potential for the geometry in Fig. \ref{linhyp} can be expressed as \cite{Ghosh}
\begin{equation}
\phi(x,y,t)=(U_0-V_0\cos{(\Omega_Tt)})\left(\frac{x^2-y^2}{2r_0^2}\right)
\end{equation}
where $r_0$ is defined as the ion-electrode distance. This is from the centre of the trap to the nearest electrode and $\Omega_T$ is the drive frequency of the applied time varying voltage. The equations of motion of the ion due to the above potential are then given by \cite{Ghosh}
\begin{equation}\label{eomx}
\frac{d^2x}{dt^2}=-\frac{e}{m}\frac{\partial\phi(x,y,t)}{\partial x}=-\frac{e}{m r_0^2}(U_0-V_0\cos{\Omega_Tt})x\\
\end{equation}
\begin{equation}\label{eomy}
\frac{d^2y}{dt^2}=-\frac{e}{m}\frac{\partial\phi(x,y,t)}{\partial y}=\frac{e}{m r_0^2}(U_0-V_0\cos{\Omega_Tt})y\\
\end{equation}
\begin{equation}
\frac{d^2z}{dt^2}=0
\end{equation}
Later it will be shown that the confinement in the $z$-axis will be produced from the addition of a static potential. Making the following substitution
\begin{equation*}
a_x=-a_y=\frac{4eU_0}{mr_0^2\Omega_T^2}, \qquad q_x=-q_y=\frac{2eV_0}{mr_0^2\Omega_T^2}, \qquad \zeta=\Omega_Tt/2
\end{equation*}
equations \ref{eomx} and \ref{eomy} can be written in the form of the Mathieu equation.
\begin{equation}\label{Mathieu}
\frac{d^2i}{d\zeta^2}+(a_i-2q_i\cos{2\zeta})i=0, \qquad i=[x,y]
\end{equation}
The general Mathieu equation given by equation \ref{Mathieu} is periodic due to the $2q_i\cos{2\zeta}$ term. The Floquet theorem \cite{Abramowitz} can be used as a method for obtaining a solution. Stability regions for certain values of the $a$ and $q$ parameters exist in which the ion motion is stable. By considering the overlap of both the stability regions for the $x$ and $y$-axes of the trap \cite{Ghosh,Horvath}, the parameter region where stable trapping can be accomplished is obtained. For the case when $a=0$, $q\ll1$ then the motion of the ion in the $x$-axis can be described as follows,
\begin{equation}
x(t) = x_0\cos(\omega_xt)\left[1+\frac{q_x}{2}\cos(\Omega_Tt)\right]
\end{equation}
with the equation of motion in the $y$-axis of the same form. The motion of the ion is composed of secular motion $\omega_x$ (high amplitude slow frequency) and micromotion at the drive frequency $\Omega_T$ (small amplitude high frequency).\\

The second form to calculate motion is the pseudopotential approximation \cite{Dehmelt}. This considers the time averaged force experienced by the ion in an inhomogeneous field. With an rf voltage of $V_0\cos{\Omega_Tt}$ applied to the trap a solution for the pseudopotential approximation is given by \cite{Dehmelt}
\begin{equation}\label{pseudo}
\psi(x,y,z)=\frac{e^{2}}{4m\Omega_{T}^{2}}|\nabla V(x,y,z)|^2
\end{equation}
where m is the mass of the ion, $\nabla V(x,y,z)$ is the gradient of the potential. The motion of the ion in a rf potential can be described just by the secular motion in the limit where $q_i/2\equiv\sqrt{2}\omega_i/\Omega_T\ll1$. The secular frequency of the ion is given by \cite{Madsen}.
\begin{equation}
\omega_{i}^2(x,y,z)=\frac{e^2}{4m^{2}\Omega_{T}^{2}}\frac{\partial^2}{\partial x^2}(|\nabla V(x,y,z)|^2)
\end{equation}
The pseudopotential approximation provides a means to treat the rf potential in terms of electrostatics only, leading to simpler analysis of electrode geometries.\\

Micromotion can be divided into intrinsic and extrinsic micromotion. Intrinsic micromotion refers to the driven motion of the ion when displaced from the rf nil position due to the secular oscillation within the trap. Extrinsic micromotion describes an offset of the ion's position from the rf nil from stray electric fields, this can be due to imperfections of the symmetry in the construction of the trap electrodes or the build up of charge on dielectric surfaces. Micromotion can cause a problem with the widening of atomic transition linewidth, second-order Doppler shifts and reduced lifetimes without cooling \cite{Berkeland}. It is therefore important when designing ion traps that compensation of stray electric fields can occur in all directions of motion. Another important factor is the occurrence of a possible phase difference $\varphi$ between the rf voltages on different rf electrodes within the ion trap. This will result in micromotion that cannot be compensated for. A phase difference of $\varphi=1^{\circ}$ can lead to an increase in the equivalent temperature for the kinetic energy due to the excess micromotion of 0.41 K \cite{Berkeland}, well above the Doppler limit of a few milli-kelvin. \\

The trap depth of an ion trap is the potential difference between pseudopotential at the minimum of the ion trap and the lowest turning point of the potential well. For hyperbolic geometries this is at the surface of the electrodes, for linear geometries (see section \ref{linear}) it can be obtained through electric field simulations. Higher trap depths are preferable as they allow the ion to remain trapped longer without cooling. Typical trap depths are on the order of a few eV. The speed of optical qubit gates for quantum information processing \cite{Steane2} and shuttling within arrays \cite{Hucul} is dependent on the secular frequency of the ion trap. Secular frequencies and trap depth are a function of applied voltage, drive frequency $\Omega_T$, the mass of the ion m and the particular geometry (particular the ion - electrode distance). Since the variation of the drive frequency is limited by the stability parameters, it is important to achieve large maximal rf voltages for the design of microfabricated ion traps, which is typically limited by bulk breakdown and surface flashover (see Section \ref{ElecChara}). It is also important to note that the secular frequency also increases whilst scaling down trap dimensions for a given applied voltage allowing for large secular frequencies at relatively small applied voltages.

\subsection{Motional and internal states of the ion.}
Single ions can be considered to be trapped within a three-dimensional harmonic well with the three directions of motion uncoupled. Considering the motion of the ion along one of the axes, the Hamiltonian describing this model can be represented as
\begin{equation}
\textit{H}=\hbar \omega\left(a^{\dag}a+\frac{1}{2}\right)
\end{equation}
with $\omega$ the secular frequency, $a^{\dag}$ and a the raising and lowering operators respectively, these operators have the following properties $a^{\dag}|n\rangle =\sqrt{n+1}|n+1\rangle$, $a|n\rangle =\sqrt{n}|n-1\rangle$. When an ion moves up one motional level; it is said to have gained one motional quantum of kinetic energy. For most quantum gates with trapped ions, the ion must reside in within the Lambe Dicke regime. This is where the ion's wave function spread is much less than the optical wavelength of the photons interacting with the ion. The original proposed gates \cite{CiracandZoller} required the ion to be in the ground state motional energy level, but more robust schemes \cite{SorensenMolmer} do not have such stringent requirements anymore.\\

Another requirement for many quantum technology application is the availability of a two-level system for the qubit to be represented, such that the ion's internal states can be used for encoding. The qubit can then be initialised into the state $|1\rangle$, $|0\rangle$ or a superposition of both. Typical ion species used are hydrogenic ions, which are left with one orbiting electron in the outer shell and similar structure to hydrogen once ionised. These have the simplest lower level energy diagrams. Candidates for ions to be used as qubits can be subdivided into two categories. Hyperfine qubits, $^{171}$Yb$^+$, $^{43}$Ca$^+$, $^{9}$Be$^+$, $^{111}$Cd$^+$, $^{25}$Mg$^+$ use the hyperfine levels of the ground state and have lifetimes on the order of thousands of years, whilst optical qubits, $^{40}$Ca$^+$, $^{88}$Sr$^+$, $^{172}$Yb$^+$ use a ground state and a metastable state as the two level system. These metastable states typically have lifetimes on the order of seconds and are connected via optical transitions to the other qubit state.

\subsection{Laser cooling}\label{Lasercool}
For most applications, the ion has to be cooled to a state of sufficiently low motional quanta, which can be achieved via laser cooling. For a two level system, when a laser field with a frequency equivalent to the spacing between the two energy levels is applied to the ion, photons will be absorbed resulting in a momentum ``kick" onto the ion. The photon is then spontaneously emitted which leads to another momentum kick onto the ion in a completely random direction so the net effect of many photon emissions averages to zero. Due to the motion of the ion within the harmonic potential the laser frequency will undergo a Doppler shift. By red detuning (lower frequency) the laser frequency by $\delta$ from resonance, Doppler cooling can be achieved. When the ion moves towards the laser, the ion will experience a Doppler shift towards the resonant transition frequency and more scattering events will occur with the net momentum transfer slowing the ion down. Less scattering events will occur when travelling away from the Doppler shifted laser creating a net cooling of the ion's motion.
Doppler cooling can typically only achieve an average motional energy $\bar{n} > 1$. In order to cool to the ground state of motion, resolved sideband cooling can be utilized. This can be achieved with stimulated Raman transitions \cite{Monroe,King}.\\

For effective cooling of the ion, the $\vec{k}$-vector of the laser needs a component in all three directions of uncoupled motion. These directions depend on the trap potential and they are called the principal axes. For a convenient choice of directions for the laser beam, principal axes can be rotated by an angle $\theta$ by application of appropriate voltages \cite{Madsen,Allcock} or asymmetries in the geometry about the ion's position \cite{Britton2,Amini,Nizamani}. The angle of rotation of the principal axes can be obtained through the Hessian matrix of the electric field. This angle describes a linear transformation of the electric potential that eliminates any cross terms between the axes creating uncoupled equations of motion. The eigenvectors of the matrix signify the direction of the principal axes.\\

\subsection{Operation of microfabricated ion traps}
In order to successfully operate a microfabricated ion trap a certain experimental infrastructure needs to be in place, from the ultra high vacuum (UHV) system apparatus to the radio-frequency source. A description of experimental considerations for the operation of microfabricated ion traps was given by McLoughlin et al. \cite{McLoughlin}. For long storage times of trapped ions and performing gate operations, the collision with background particles must not be a limiting factor. Ion traps are therefore typically operated under ultra high vacuum (UHV) (pressures of $10^{-9}-10^{-12}$ mbar). The materials used need to be chosen carefully such that outgassing does not pose a problem. The materials used for different trap designs are discussed in more detail within section \ref{FabProcess}.\\

To generate the high rf voltage ($\sim$100-1000V) a resonator \cite{Siverns} is commonly used. Typical resonator designs include helical and coaxial resonators. The advantage of using a resonator, is that it provides a frequency source with a narrow bandpass, defined by the quality factor Q of the combined resonator - ion trap circuit. This provides a means of filtering out frequencies that couple to the motion of the ion leading to motional heating of the trapped ion. A resonator also fulfills the function of impedance matching the frequency source to the ion trap. The total resistance and capacitance of the trap lowers the Q factor. It is important to minimize the resistance and capacitance of the ion trap array if a high Q value is desired.\\

\begin{figure}
\begin{center}
\resizebox*{10cm}{!}{\includegraphics{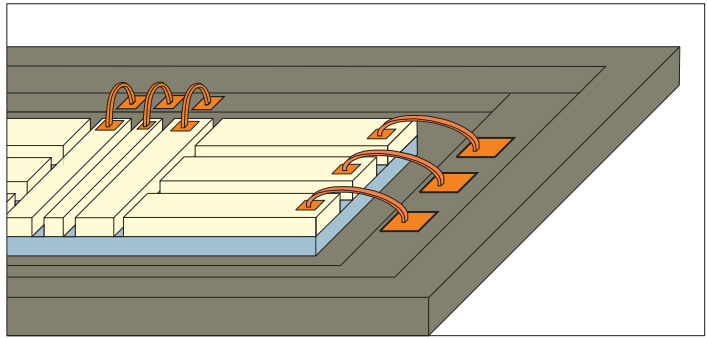}}%
\caption{Wire bonding a microfabrciated chip to a chip carrier providing external electrical connections.}  
\label{bondpads}
\end{center}
\end{figure}

To provide electrical connections, ion traps are typically mounted on a chip carrier with electrical connections provided by wire bonding individual electrodes to an associate connection on the chip carrier. Bond pads on the chip are used to provide a surface in which the wire can be connect to, see Fig. \ref{bondpads}. The pins of the chip carrier are connected to wires which pass to external voltage supplies outside the vacuum system.\\

The loading of atomic ions within Paul traps is performed utilizing a beam of neutral atoms typically originating from an atomic oven consisting of a resistively heated metallic tube filled with the appropriate atomic species or its oxide. The atomic flux is directed to the trapping region where atoms can be ionised via electron bombardment or more commonly by photoionisation, see for example refs. \cite{McLoughlin,Deslauriers2}. The latter has the advantage of faster loading rates requiring lower neutral atom pressures and results in less charge build up resulting from electron bombardment. For asymmetric traps in which all the electrodes lie in the same plane, see Fig. \ref{lineartraps}, a hole within the electrode structure can be used for the atomic flux to pass through the trap structure, this is defined as backside loading \cite{Britton2}. The motivation behind this method is to reduce the coating of the electrodes and more importantly reduce coating of the notches between the electrodes from the atomic beam reducing charge build up and the possibility of shorting between electrodes. However, the atomic flux can also be directed parallel to the surface in an asymmetric ion trap due to the low atomic flux required for photoionisation loading.

\section{Linear ion traps}\label{linear}
The previously mentioned ideal linear hyperbolic trap only provides confinement within the radial directions and does not allow for optical access. By modifying the geometry as depicted in Fig. \ref{lineartraps} linear ion traps are created. To create an effective static potential for the confinement in the axial ($z$-axis) direction, the associate electrodes are segmented. This allows for the creation of a saddle potential and when superimposed onto rf pesduopotenial provides trapping in three dimensions. By selecting the appropriate amplitudes for the rf and static potentials such that the radial secular frequencies $\omega_x,\omega_y$ are significantly larger than the axial frequency $\omega_z$, multiple ions will form a linear chain along the $z$-axis. The motion of the ion near the centre of the trap can be considered to be harmonic as a very good approximation. The radial secular frequency of a linear trap is on the order of that of a hyperbolic ion trap of same ion - electrode distance but different by a geometric factor $\eta$ \cite{Madsen}.
\subsection{Linear ion trap geometries}
\begin{figure}[htp]
\begin{center}
\subfigure[]{
\resizebox*{5cm}{!}{\includegraphics{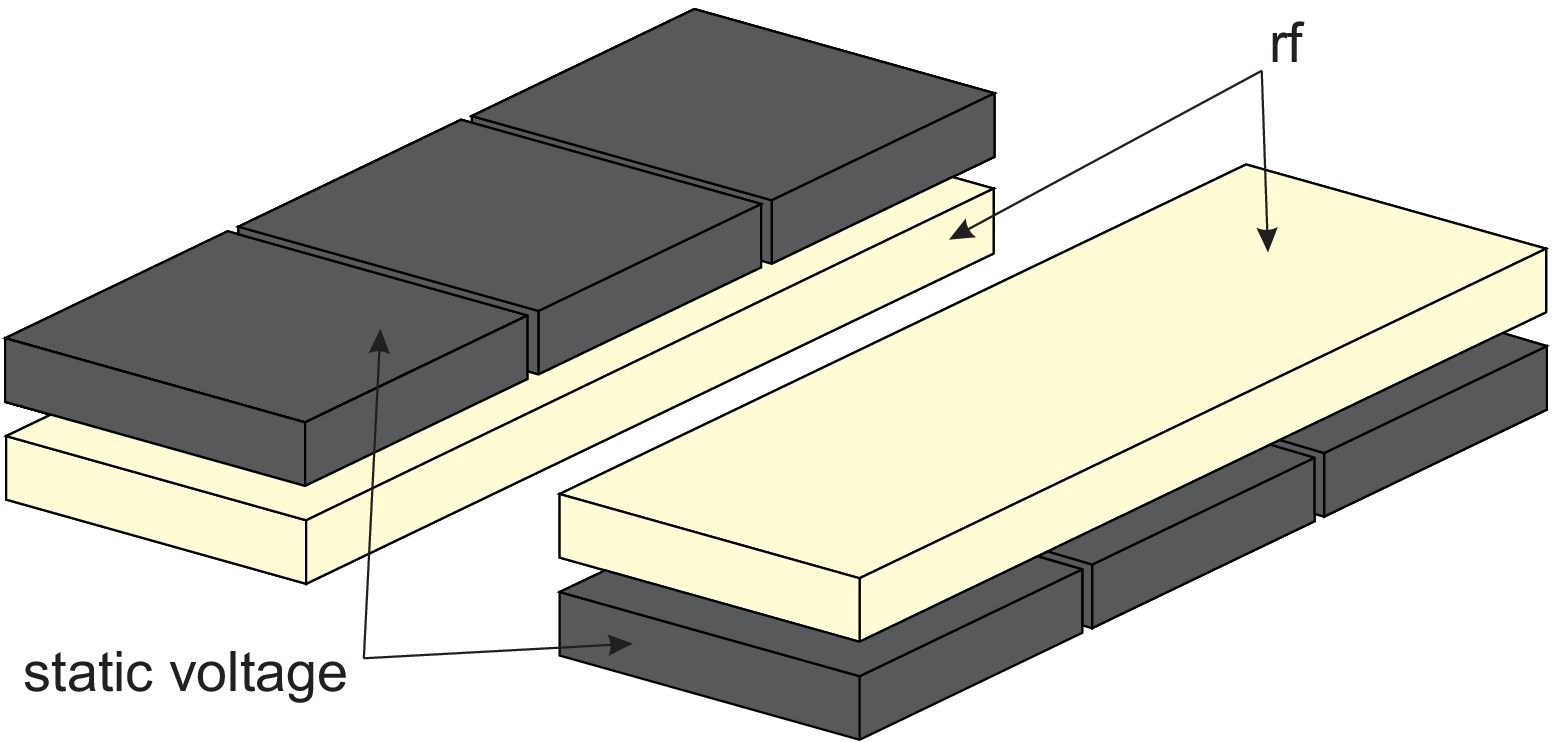}}}%
\subfigure[]{
\resizebox*{5cm}{!}{\includegraphics{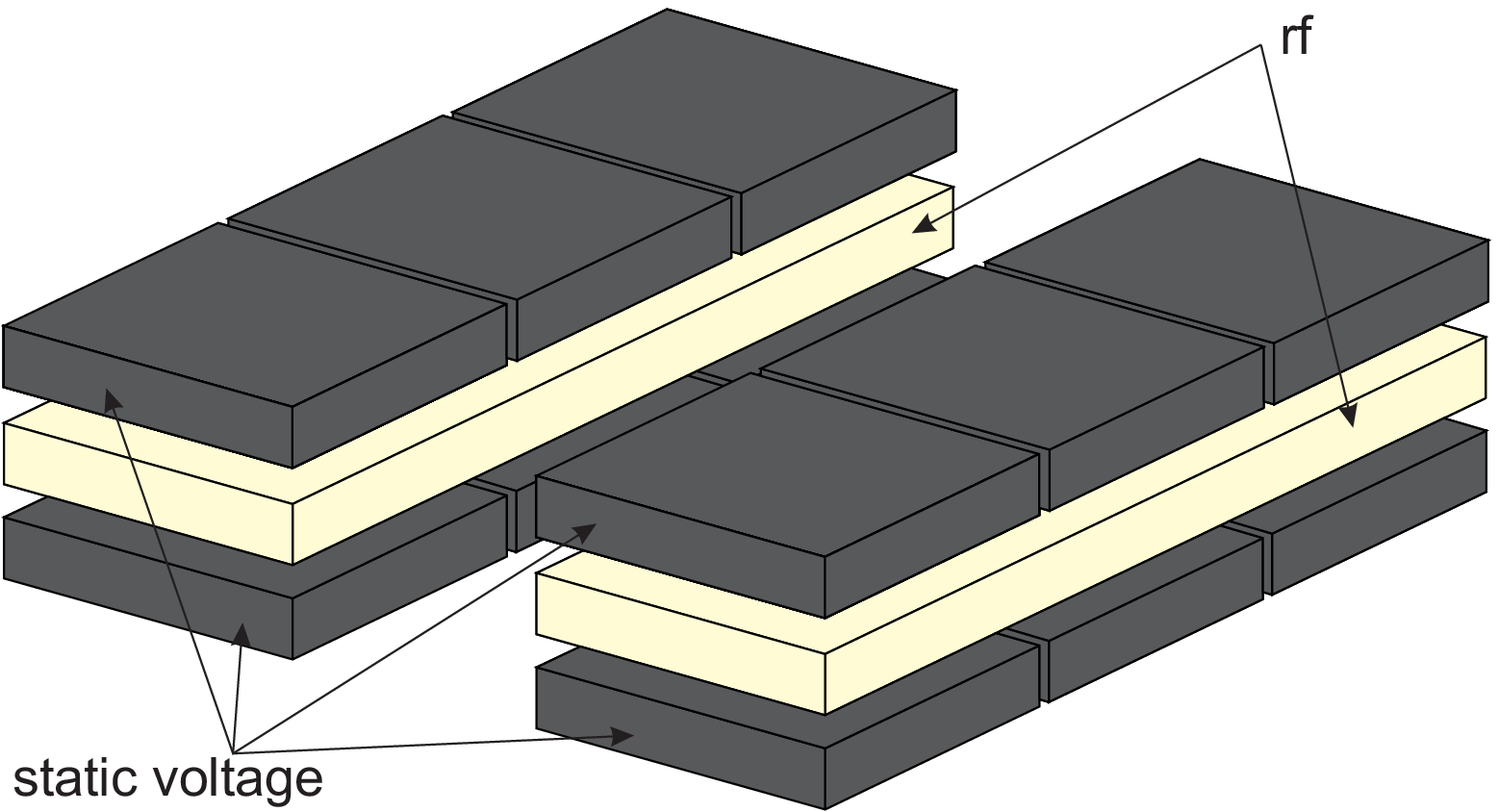}}}%
\subfigure[]{
\resizebox*{5cm}{!}{\includegraphics{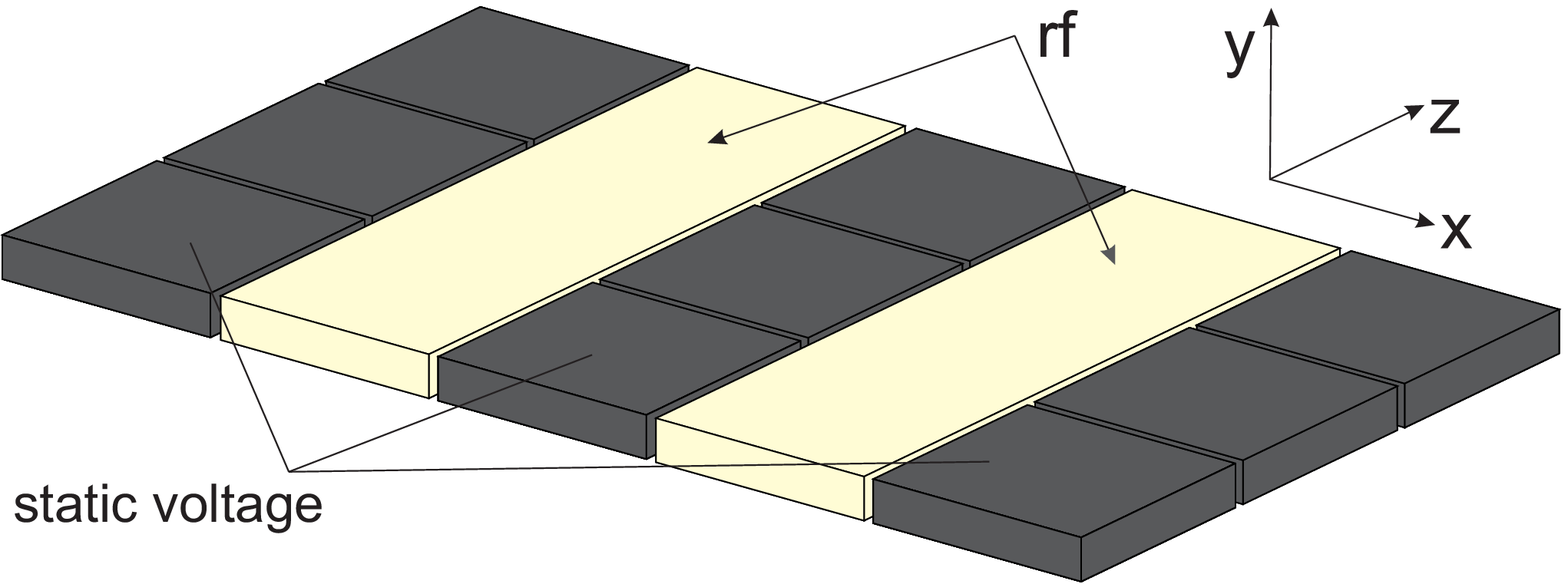}}}%
\caption{Different linear trap geometries. (a) A two-layer design in which the rf electrodes (yellow) are diagonally opposite and the dc electrodes (grey) are segmented.
(b) A three-layer design in which the rf electrodes are surrounded by the dc electrodes. (c) A five-wire asymmetric design where all the electrodes lie in the same plane. } %
\label{lineartraps}
\end{center}
\end{figure}

Linear ion trap geometries can be realised in a symmetric or asymmetric design as depicted in Fig. \ref{lineartraps}.
In symmetric designs the ions are trapped between the electrodes, as shown for two- and three-layer designs in Fig. \ref{lineartraps} (a) and (b) respectively. These types of designs offer higher trap depths and secular frequencies compared to asymmetric traps of the same trap parameters. Two-layer designs offer the highest secular frequencies and trap depths whilst three-layer designs offer more control of the ions position for micromotion compensation and shuttling.\\

\begin{figure}[htp]
\begin{center}
\resizebox*{8cm}{!}{\includegraphics{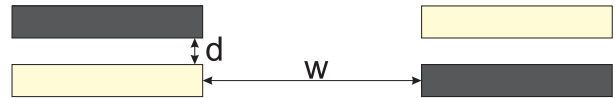}}%
\caption{for two-layer traps the aspect ratio is defined as w/d}  
\label{AR}
\end{center}
\end{figure}

The aspect ratio for symmetric designs is defined as the ratio of the separation between the two sets of electrodes w and the separation of the layers d depicted in Fig. \ref{AR}. As the aspect ratio rises, the geometric efficiency factor $\eta$ decreases and approaches asymptotically $1/\pi$ for two-layer designs. \cite{Madsen}.
Another advantage of symmetric traps is more freedom of optical laser access, allowing laser beams to enter the trapping zone in various angles. In asymmetric trap structures the laser beams typically have to enter the trapping zone parallel to the trap surface. Asymmetric designs offer the possibility of simpler fabrication processes. Buried wires \cite{Amini} and vertical interconnects can provide electrical connections to electrodes which cannot be connected via surface pathways. Traps depths are typically smaller than for symmetric ion traps, therefore higher voltages need to be applied to obtain the same trap depth and secular frequencies of an equivalent symmetric ion trap. The widths of the individual electrodes can be optimised to maximise trap depth \cite{Nizamani}.\\

\begin{figure}[htp]
\begin{center}
\subfigure[][]{
\resizebox*{5cm}{!}{\includegraphics{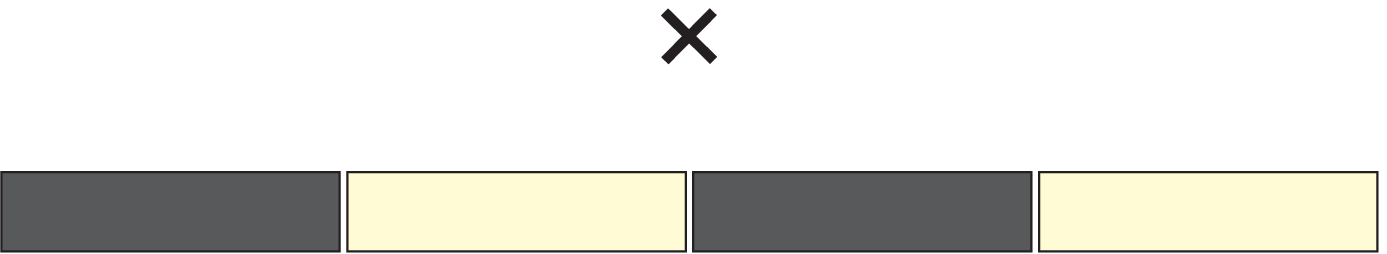}}}%
\\
\subfigure[]{
\resizebox*{5cm}{!}{\includegraphics{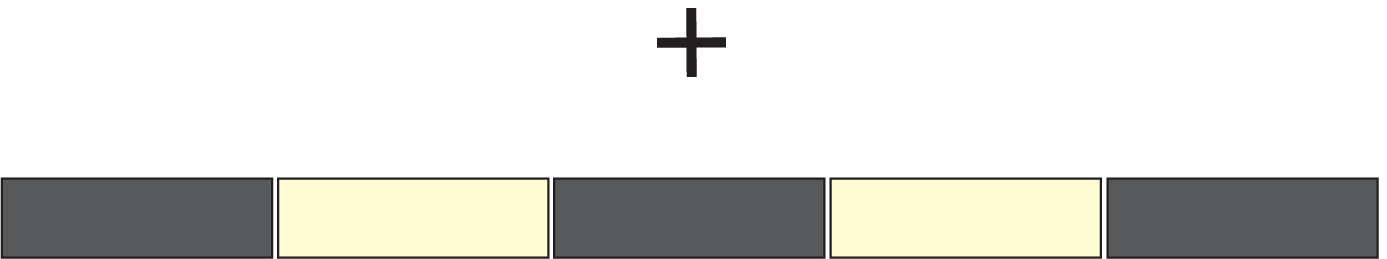}}}%
\\
\subfigure[]{
\resizebox*{5cm}{!}{\includegraphics{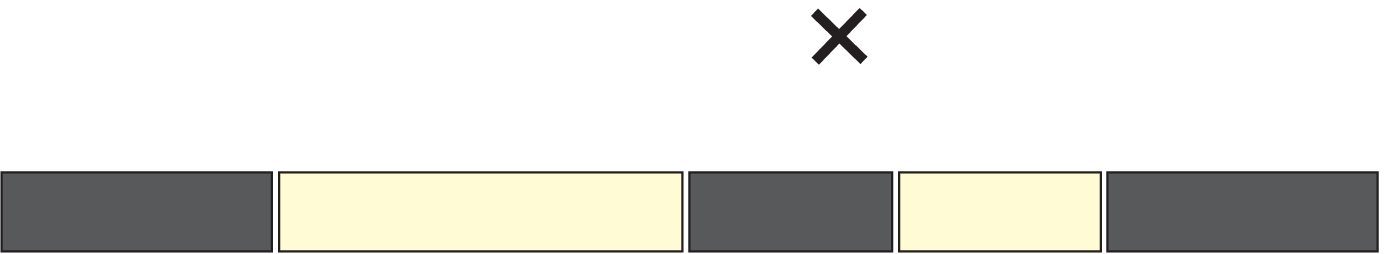}}}%
\\
\subfigure{
\resizebox{2cm}{!}{\includegraphics{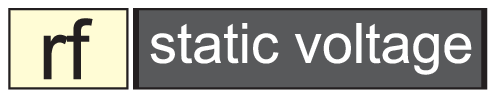}}}%
\caption{Cross-section in the $x$-$y$ plane of the different types of asymmetric designs, (a) Four-wire design in which the principal axes are naturally non-perpendicular with respect to the plane of the electrodes.
(b) Five-wire design where the electrodes are symmetric and one principal axis is perpendicular to the surface.
(c) Five-wire design with different widths rf electrodes, this allows the principal axes to be rotated.} %
\label{asymmtraps}
\end{center}
\end{figure}
In order to successfully cool ions, Doppler cooling needs to occur along all three principal axes therefore the $\vec{k}$-vector of the laser needs to have a component along all the principal axes. Due to the limitation of the laser running parallel to the surface it is important that all principal axes have a component along the $\vec{k}$-vector of the Doppler cooling laser beam. Five-wire designs (Fig. \ref{asymmtraps}(b) and (c)) have a static voltage electrode below the ions position, surrounded by rf electrodes and additional static voltage electrodes. With the rf electrodes of equal width (Fig. \ref{asymmtraps}(b)) one of the principal axes is perpendicular to the surface of the trap. However, it can be rotated via utilizing two rf electrodes of different width (Fig. \ref{asymmtraps}(c)) or via splitting the central electrodes \cite{Allcock}. A four-wire design shown in Fig. \ref{asymmtraps}(a) has the principal axes naturally rotated but the ion is in direct sight of the dielectric layer below since the ion is located exactly above the trench separating two electrodes. Deep trenches have been implemented \cite{Britton,Britton2} to reduce the effect of exposed dielectrics.

\subsection{From linear ion traps to arrays}
For ions stored in microfabricated ion traps to become viable for quantum information processing, thousands or even millions of ions need to be stored and interact with each other. This likely requires a number of individual trapping regions that is on the same order as the number of ions and furthermore, the ability for the ions to interact with each other so that quantum information can be exchanged. This could be achieved via arrays of trapping zones that are connected via junctions. To scale up to such an array requires fabrication methods that are capable of producing large scale arrays without requiring an unreasonable overhead in fabrication difficulty. This makes some fabrication methods more viable for scalability than other techniques. An overview of different fabrication methods is discussed in more detail within section \ref{FabProcess}.\\

The transport of ions through junctions has first been demonstrated within a three-layer symmetric design \cite{Hensinger} and later near-adiabatic in a two-layer symmetric trap array \cite{Blakestad}. Both ion trap arrays were made from laser machined alumina substrates incorporating mechanical alignment. The necessity of mechanical alignment and laser machining limit the opportunity to scale up to much larger numbers of electrodes making other microfabrication methods more suitable in the long term. Transport through an asymmetric ion trap junction was then demonstrated by Amini et al. \cite{Amini}, however, this non adiabatic transport required continuous laser cooling. Wesenberg carried out a theoretical study \cite{Wesenberg2} how one can implement optimal ion trap array intersections. Splatt et al. demonstrated reordering of ions within a linear trap \cite{Splatt}.\\
\begin{figure}[htp]
\begin{center}
\subfigure[]{
\resizebox*{5cm}{!}{\includegraphics{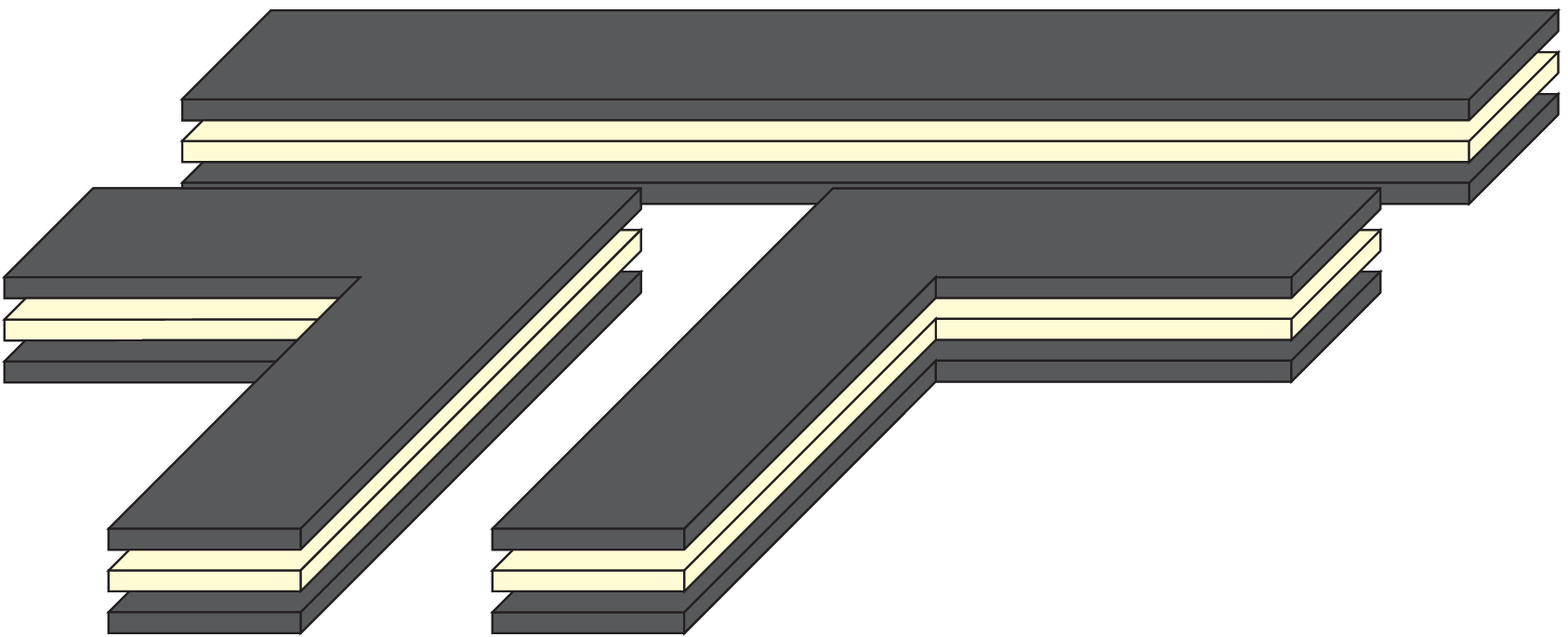}}}%
\subfigure[]{
\resizebox*{5cm}{!}{\includegraphics{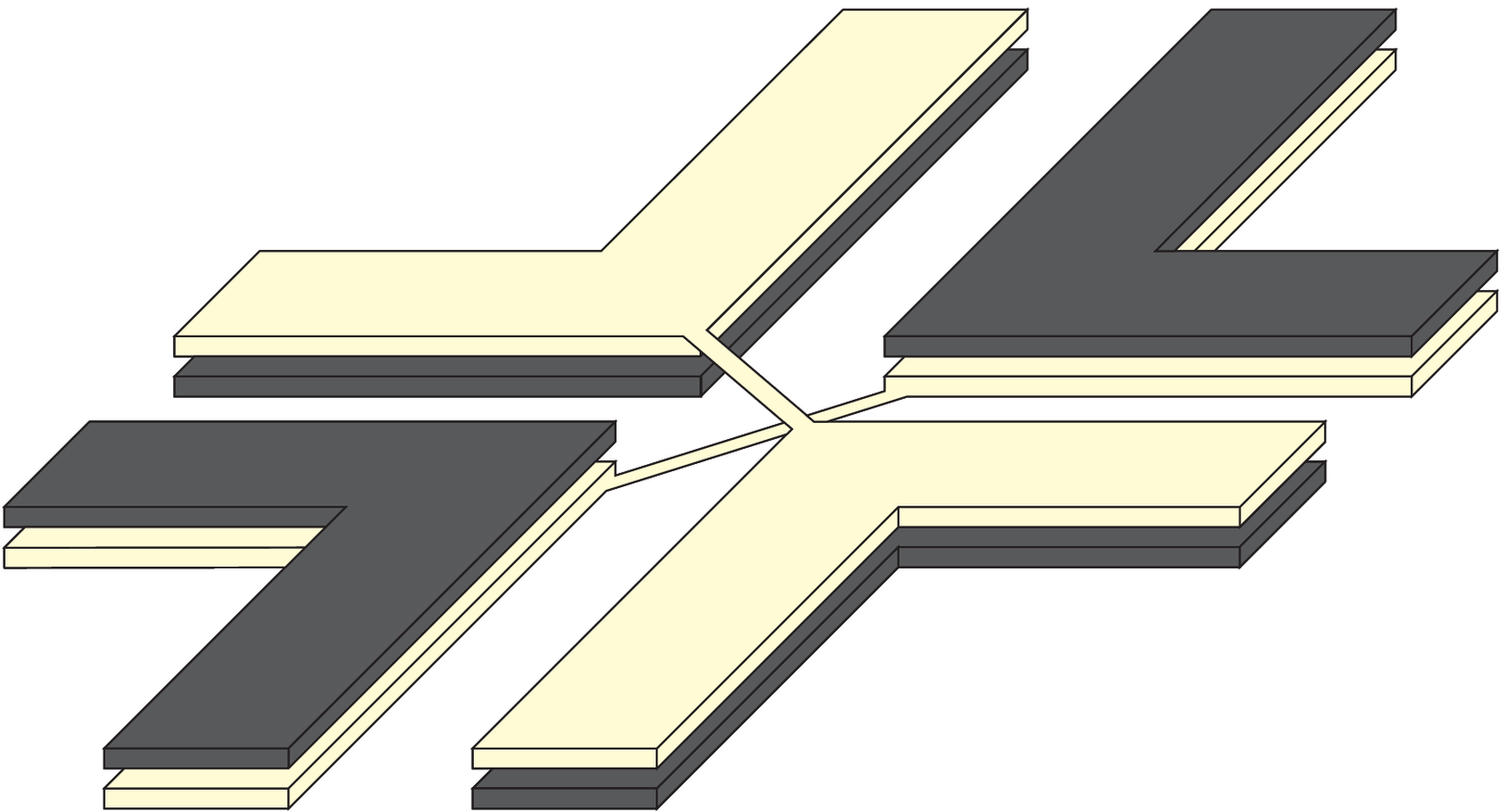}}}%
\subfigure[]{
\resizebox*{5cm}{!}{\includegraphics{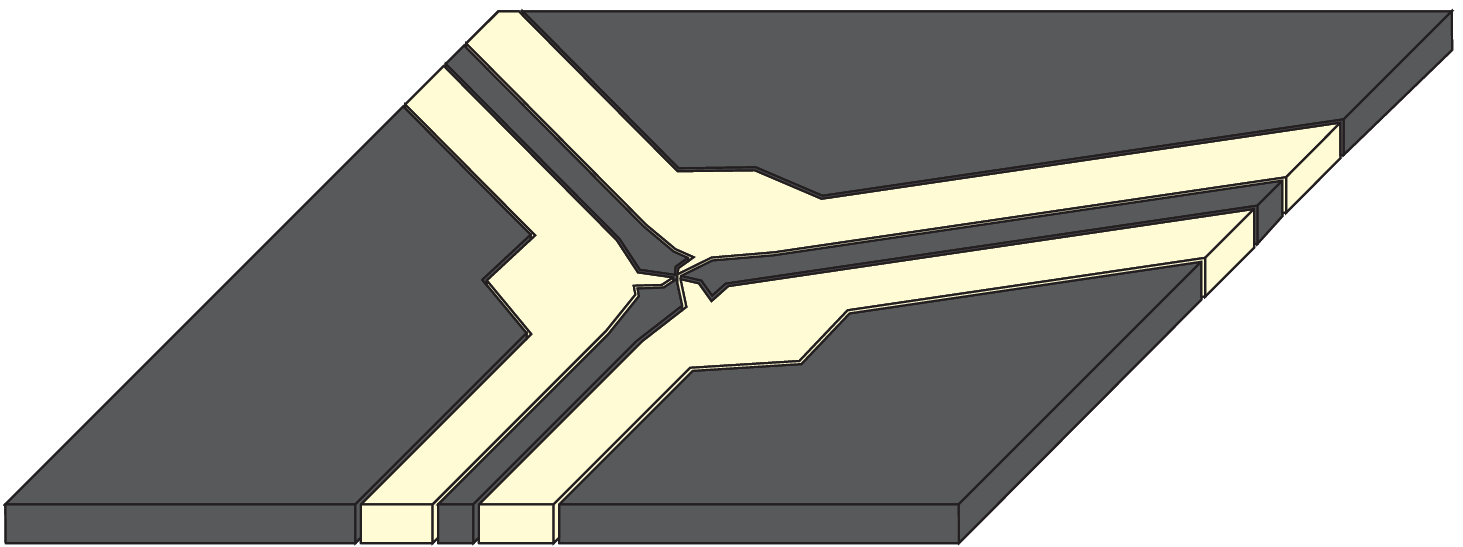}}}%
\caption{Junctions that have been used to successfully shuttle ions. The yellow parts represent the rf electrodes. No segmentation of the static voltage electrodes (grey) is shown.
(a) T-junction design \cite{Hensinger} where corner shuttling and swapping of two ions were demonstrated for the first time. While the transport was reliable the ion gained a significant amount of kinetic energy during a corner-turning operation.
(b) A two-layer X-junction \cite{Blakestad} was used to demonstrate highly reliable transport through a junction with a kinetic energy gain of only a few motional quanta.
(c) A Y-junction \cite{Amini} was used to demonstrate transport through an asymmetric junction design, however, requiring continuous laser cooling during the shuttling process.} %
\label{junctions}
\end{center}
\end{figure}

\section{Simulating the electric potentials of ion trap arrays}\label{simu}
Accurate simulations of the electric potentials are important for determining trap depths, secular frequencies and to simulate adiabatic transport including the separation of multiple ions and shuttling through corners \cite{Hucul,Reichle}. Various methods can be used in determining the electric potentials from the trap electrodes, with both analytical and numerical methods available. Numerical simulations using the finite element method (FEM) and the boundary element method (BEM) \cite{Hucul,Singer} provide means to obtain the full 3D potential of the trap array. FEM works by dividing the region of interest into a mesh of nodes and vertices, an iterative process then finds a solution which connects the nodes whilst satisfying the boundary conditions and a potential can be found for each node. BEM starts with the integral equation formulation of Laplace's equation resulting in only surface integrals being non-zero in an empty ion trap. Due to BEM solving surface integrals, this is a dimensional order less than FEM thus providing a more efficient numerical solution than FEM \cite{Hucul}. To obtain the total potential the basis function method is used \cite{Hucul}. A basis function for a particular electrode is obtained by applying 1V to one particular electrode whilst holding the other electrodes at ground. By summing all the basis functions (with each basis function multiplied by the actual voltage for the particular electrode) the total trapping potential can be obtained.\\

For the case of asymmetric ion traps, analytical methods provide means to calculate the trapping potential at a quicker rate and scope for optimisation of the electrode structures. A Biot-Savart-like law \cite{Oliveira} can be used and is related to the Biot-Savart law for magnetic fields in which the magnetic field at a point of interest is obtained by solving the line integral of an electrical current around a closed loop. This analogy is then applied to electric fields in the case of asymmetric ion traps \cite{Wesenberg}. One limitation for these analytical methods is the fact that all the electrodes must lie in a single plane, with no gaps, which is referred to as the gapless plane approximation. House \cite{House} has obtained analytical solutions to the electrostatic potential of asymmetric ion trap geometries with the electrodes located on a single plane within a gapless plane approximation. Microfabrication typically requires gaps of a few micrometers \cite{Britton2} which need to be created to allow for different voltages on neighboring electrodes. The approximation is suggested to be reasonable for gaps much smaller than the electrode widths and studies into the effect of gapped and finite electrodes have been conducted \cite{Schmied}. However, within the junction region where electrodes can be very small and high accuracy is required, the gapless plane approximation may not necessarily be sufficient.

\section{Electrical characteristics}\label{ElecChara}

\subsection{Voltage breakdown and surface flashover}
Miniaturization of ion traps is not only limited by the increasing motional heating of the ion (see section \ref{heating}), but also by the maximum applied voltages allowed by the dielectrics and gaps separating the electrodes. Both secular frequency and trap depth depend on the applied voltage. Therefore it is important to highlight important aspects involved in electrical breakdown. Breakdown can occur either through the bulk material, a vacuum gap between electrodes, or across an insulator surface (surface flashover). There are many factors which contribute to the breakdown of a trap, from the specific dielectric material used and its deposition process, residues on insulating materials to the geometry of the electrodes itself and the frequency of the applied voltage.\\

Bulk breakdown describes the process of breakdown via the dielectric layer between two independent electrodes. An important variable which has been modelled and measured is the dielectric strength. This is the maximum field that can be applied before breakdown occurs. The breakdown voltage $V_c$ is related to the  dielectric strength for an ideal capacitor by $V_c=dE_c$, where d is the thickness of the dielectric. There have been many studies into dielectric strengths showing an inverse power law relation $E_c\propto d^{-n}$ \cite{Agarwal1,Agarwal2,MRB82,KS01,ZSZ03,B03,MPC}. The results show a typical range of values ($0.5-1$) for the scaling parameter $n$. Although decreasing the thickness will increase the dielectric strength this will not increase the breakdown voltage if the scaling parameter lies below one.\\

Surface flashover occurs over the surface of the dielectric material between two adjacent electrodes. The topic has been reviewed \cite{Miller} with studies showing a similar trend with a distance dependency on the breakdown voltage with $V_b\propto d^\alpha$ where $\alpha\approx0.5$ \cite{Pillai2,MPC}. Surface flashover usually starts from electron emission from the interface of the electrode, dielectric and vacuum known as the triple point. Imperfections at this point increase the electric field locally and will reduce the breakdown voltage. The electric field strength for surface breakdown has been measured to be a factor of 2.5 less than that for bulk breakdown of the same material, dimensions and deposition process \cite{MPC}, with thicknesses of $1-3.9\mu m$ for substrate breakdown and lengths of $5-600\mu m$ considered.\\

The range of parameters that can affect breakdown from the difference between rf and applied static voltages \cite{Stick,Pillai,Pillai2} includes the dielectric material, deposition process and the geometry of the electrodes \cite{Miller}. Therefore it is most advisable to carry out experimental tests on a particular ion trap fabrication design to determine reliable breakdown parameters. In a particular design it is very important to avoid sharp corners or similar features as they will give rise to large local electric fields at a given applied voltage.

\subsection{Power dissipation and loss tangent}\label{power}

\begin{figure}[h]
\begin{center}
\resizebox*{8cm}{!}{\includegraphics{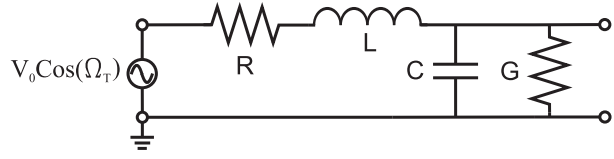}}%
\caption{The rf electrode modelled with resistance $R$, capacitance $C$, inductance $L$ and conductance $G$.}%
\label{circuitmodel}
\end{center}
\end{figure}

When scaling to large trap arrays the finite resistance $R$ and capacitance $C$ of the electrodes as well as the dielectric materials within the trap structure result in losses and therefore have to be taken into account when designing an ion trap. The power dissipation, which results from rf losses, is highly dependent on the materials used and the dimensions of the trap structure and can result in heating and destruction of trap structures. To calculate the power dissipated in a trap a simple lumped circuit model, as shown in Fig. \ref{circuitmodel} can be utilized.\\

The dielectric material insulating the electrodes cannot be considered a perfect insulator resulting in a complex permittivity $\varepsilon$, with $\varepsilon=\varepsilon'-j\varepsilon''$. Lossless parts are represented by $\varepsilon'$ and lossy parts by $\varepsilon''$. Substituting this into Amp\`ere's circuital law and by rearranging one obtains \cite{Kraus}
\begin{equation}
{\bf \nabla \times H} = \left[ \left( \sigma + \omega \epsilon ''\right) + j \omega
\epsilon '\right]{\bf E}
\end{equation}
$\sigma$ is the conductivity of the dielectric for an applied alternating current, and $\omega$ is the angular frequency of the applied field. The effective conductance is often referred to as $ \sigma' = \sigma + \omega \epsilon ''$. The ratio of the conduction to displacement current densities is commonly defined as loss tangent $\tan \delta$
\begin{equation}
\tan \delta = \frac{\sigma + \omega \epsilon ''}{\omega \epsilon '}
\end{equation}
For a good dielectric the conductivity $\sigma$ is much smaller then  $\omega \epsilon ''$ and we can then make the following approximation $\tan \delta = \epsilon '' / \epsilon '$. For a parallel plate capacitor the real and imaginary parts of the permittivity can be expressed as \cite{DAG08}
\begin{equation}
\varepsilon'=\frac{Cd}{\varepsilon_{0}A}, \qquad \varepsilon''=\frac{Gd}{\varepsilon_{0}\omega A}
\end{equation}
Substituting this into the approximated loss tangent expression, the conductance $G$ can be expressed as
\begin{equation}\label{Conductance}
G=\omega C \tan \delta
\end{equation}
Now we can calculate the power dissipated through the rf electrodes driven by frequency $\omega=\Omega_T$, with $I_{rms}=V_{rms}/Z$. The total impedance for this circuit is
\begin{equation}\label{imp}
Z=R+j\omega L+\frac{G-j\omega C}{G^2+\omega^2C^2}
\end{equation}
The second term of the impedance represents the inductance $L$ of the electrodes, which we approximate with the inductance of two parallel plates separated by a dielectric with $L= \mu l (d/w)$ \cite{Thierauf}. Assuming a dielectric of thickness $d \approx 10 \mu $m,  electrode width $w \approx 100 \mu$m, electrode length $l \approx 1 $mm, magnetic permeability of the dielectric $\mu \approx 10^{-6} \frac{H}{m}$ \cite{Howard}, the inductance can be approximated to be $L\approx 10^{-10}$H. Comparing the imaginary impedance terms $j \omega L $ and $ j \omega \frac{C}{G^2+\omega^2 C^2} $, it becomes clear that the inductance can be neglected. The average power dissipated in a lumped circuit is given by $P_d=Re(V_{rms}I^{\ast}_{rms})$ \cite{Horwitz}. Using the approximation for the total impedance $Z=R+\frac{G-j\omega c}{G^2+\omega^2C^2}$ we obtain
\begin{equation}
P_d= \frac{V_{0}^{2} R (G^{2}+C^{2}\omega^{2})^{2}} {2(C^{2} \omega^{2} + R^{2} (G^{2} +C^{2} \omega^{2})^{2})}
\end{equation}
and using equation \ref{Conductance} one obtains
\begin{equation}\label{powerdiss}
P_d=\frac{V_{0}^2\Omega_{T}^{2}C^2R(1 + \tan^2 \delta)^2}{2(1+\Omega_{T}^{2}C^2R^2(1 + \tan^2 \delta)^2)}
\end{equation}
In the limit where $\tan \delta$ and $\Omega_{T} C R \ll 1$, the dissipated power can be simplified as $P_d=\frac{1}{2}V_{0}^2\Omega_{T}^{2}C^2R$. Considering equation \ref{powerdiss}, important factors to reduce power dissipation are an electrode material with low resistivity, a low capacitance of the electrode geometry and a dielectric material with low loss tangent at typical drive frequencies ($\Omega_{T}= 10-80$MHz).\\

Values for loss tangent have been studied in the GHz range for microwave integrated circuit applications \cite{Krupka} and diode structures at kHz range \cite{DAG08,T06,B06}. Generally the loss tangent decreases with increasing frequency \cite{Karatas,Selcuk} and there has been a temperature dependence shown for specific structures \cite{T06}. Values at 1 MHz can be obtained but are dependent on the structures tested; Au/SiO$_2$/n-Si $\tan\delta\sim 0.05$ \cite{DAG08}, Au/Si$_3$N$_4$/p-Si $\tan\delta\sim 0.025$, \cite{B06}, Cr/SiO$_{1.4}$/Au $\tan\delta\sim 0.09$ \cite{T08}. The loss tangent will be dependent on the specific structure and also dependent on the doping levels. It is suggested that the loss tangent for specific materials be requested from the manufacturer or measured for the appropriate drive frequency. For optimal trap operation, careful design considerations have to be made to reducing the overall resistance and capacitance of the trap structure minimising the dissipated power.

\section{Fabrication Processes}\label{FabProcess}

Using the information given in the previous sections about ion trap geometries, materials and electrical characteristics common microfabrication processes can be discussed. One of the main criteria for choice of fabrication process is the compatibility with a desired electrode geometry. Asymmetric and symmetric geometries result in different requirements and therefore we will discuss them separately. First, process designs for asymmetric traps will be discussed followed by symmetric ion trap designs and universally compatible processes. The compatibility of a process with discussed materials, geometries will be highlighted. Structural characteristics and limitations of a process will be explained and solutions will then be given. As the exact fabrication steps depend on materials used and available equipment, the process sequences will be discussed as generally possible. However, to give the reader a guideline of possible choices, material and process step details of published ion traps will be given in brackets. The discussion will start with a simple process that can be performed in a wide variety of laboratories and is offered by many commercial suppliers.

\subsection{Printed Circuit Board (PCB)}

The first discussed fabrication technique used for ion trap microfabrication is the widely available Printed Circuit Boards (PCB) process. This technology is commonly used to create electrical circuits for a wide variety of devices and does not require cleanroom technology. For ion traps generally a monolithic single or two layer PCB process is used. Monolithic processes rely on the fabrication of ion trap structures by adding to or subtracting material from one component. In contrast to this, wafer bonding, mechanical mounting or other assembly techniques are used to fabricate traps from pre-structured, potentially monolithic parts in a non-monolithic process. When combined with cleanroom fabrication such a process can be used for very precise and large scale structures, however, mechanical alignment remains an issue.\\

PCB processes commonly allow for minimal structure sizes of approximately 100$\mu$m \cite{Kenneth} and slots milled into the substrate of 500$\mu$m width. This limits the uses of this technique for large and complex designs but also results in a less complex and easier accessible process. Smaller features are possible with special equipment, which is not widely available. The following section will give a general introduction into the PCB processes used to manufacture ion traps.\\

\begin{figure}[h]
\begin{center}
\resizebox*{12cm}{!}{\includegraphics{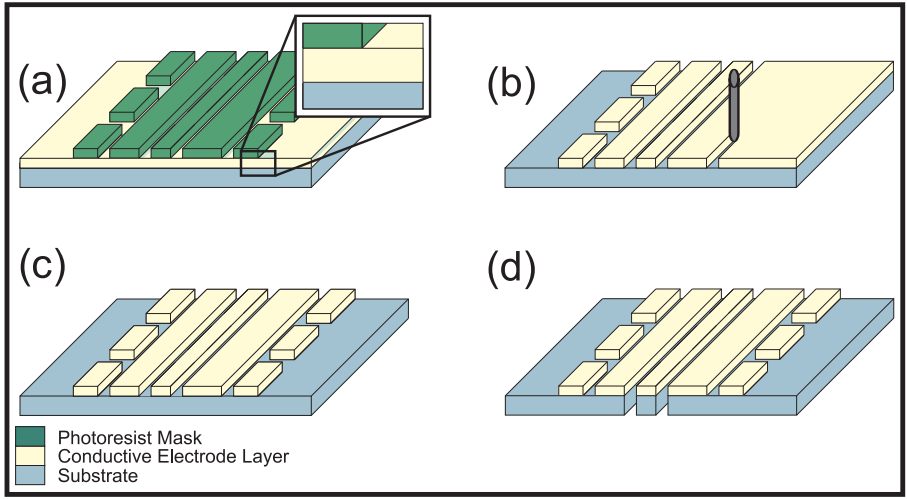}}%
\caption{Overview of several PCB fabrication processes. (a) Deposition and exposure  of the resist used as a mask in later etch steps. (b) Removal of the top copper layer using mechanical milling.
(c) Removal of the copper layer using a chemical wet etch and later removal of the deposited mask. (d) Mechanical removal of the substrate underneath the trapping zone \cite{Pearson}.}%
\label{PCB}
\end{center}
\end{figure}

This process is based on removing material instead of depositing materials as used in most cleanroom processes. Therefore the actual process sequence starts by selecting a suitable metal coated PCB substrate. The substrate has to exhibit the characteristics needed for ion trapping, ultra high vacuum (UHV) compatibility, low rf loss tangent and high breakdown voltage (commercially available high-frequency (hf) Rogers 4350B used in ref. \cite{Kenneth}). One side of the PCB substrates is generally pre-coated with a copper layer by the manufacturer, which is partially removed in the process to form the trap structures as shown in Fig. \ref{PCB} (c). Possible techniques to do this are mechanical milling shown in Fig. \ref{PCB} (b) or chemically wet etching using a patterned mask. The mask can either be printed directly onto the copper layer or photolithography can be used to pattern photo resist as shown in Fig. \ref{PCB} (a). To reduce exposed dielectrics and prevent shorting from material deposited in the trapping process slots can be milled into the substrate underneath the trapping zones as shown in Fig. \ref{PCB} (d).\\

The fabricated electrodes form one plane and make the process unusable for symmetric ion traps (SIT). Not only are the electrodes located in one plane they also sit directly on the substrate, which makes a low rf loss tangent substrate necessary to minimize energy dissipation from the rf rails into the substrate. As explained in Section \ref{power} high power dissipation leads to an increase of the temperature of the ion trap structure and also reduces the quality factor of the loaded resonator. PCB substrates intended for hf devices generally exhibit a low rf loss tangent (values of tan$\delta= 0.0031$ are typical \cite{Chenard}) and if UHV compatible can be used for ion traps. Other factors reducing the quality factor are the resistances and capacitances of the electrodes. When slots are milled into the substrate exposed dielectrics underneath the ion can be avoided despite the electrode structures being formed directly on the substrate. The otherwise exposed dielectrics would lead to charge built up underneath the ion resulting in stray fields pushing the ion out of the rf nil and leading to micromotion. By relying on widely available fabrication equipment this technique enabled the realization of several ion trap designs with micrometer-scale structures, published in \cite{Splatt,Harlander,LeibrandtPCB,Kenneth,Pearson}, without the need of cleanroom techniques.\\

The discussed single layer PCB process can be used for electrode geometries including y-junctions and linear sections. More complex topologies with isolated electrodes and buried wires require a more complex two-layer process. To address limitations caused by the minimal structure separations and sizes allowing for ion traps with higher electrode and trapping-zone densities a process based on common cleanroom technology will be discussed next.

\subsection{Conductive Structures on Substrate (CSS)} \label{CSSPro}


Common cleanroom fabrication techniques like high resolution photolithography, isotropic and anisotropic etching, deposition, electroplating and epitaxial growth allow a very precise fabrication of large scale structures. The monolithic technique we discuss here is based on a ``conductive structures on substrate" process and was used to fabricate the first microfabricated asymmetric ion trap (AIT) \cite{Seidelin}. Electrode structures separated by only 5$\mu$m \cite{Allcock} can be achieved with this process and smaller structures are not limited by the process but flashover and bulk breakdown voltage of the used materials. The structures are formed by means of deposition and electroplating of conductive material onto a substrate instead of removing precoated material. This makes it possible to use a much wider variety of materials in a low number of process steps. Several variations or additions were published \cite{Seidelin,Wang,Wang2,Labaz,StickThesis} to adjust the process for optimal results with desired materials and structures.\\

\begin{figure}[h]
\begin{center}
\resizebox*{12cm}{!}{\includegraphics{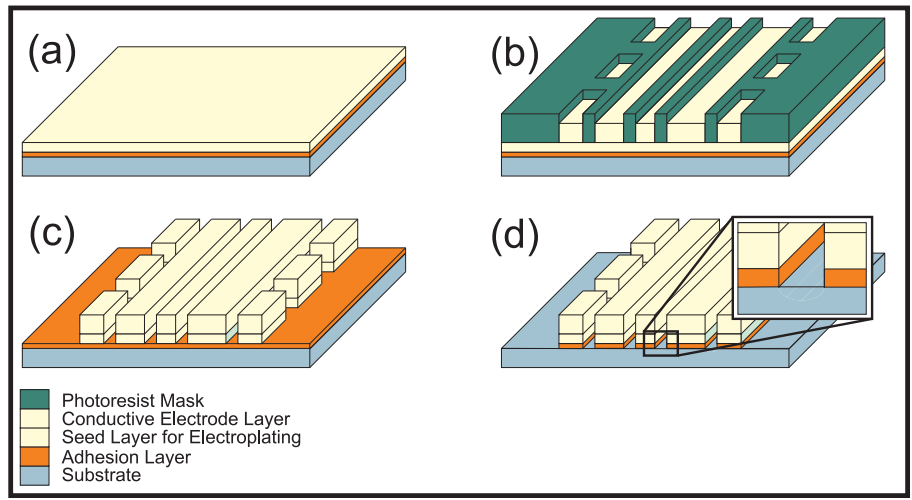}}%
\caption{Fabrication sequence of the standard ``Conductive Structures on Substrate" process.  (a) The sequence starts with the deposition of an adhesion and seed layer.
(b) An electroplating step creating the electrodes using a patterned mask. (c) Removal of the mask followed by a first etch step removing the seed layer. d) Second etch step removing the adhesion layer.}%
\label{CSS}
\end{center}
\end{figure}

First the standard process will be presented, which starts by coating the entire substrate (for example polished fused quartz as used in \cite{Seidelin}) with a metal layer working as a seed layer (0.1$\mu$m copper), necessary for a following electroplating step. Commonly an adhesion layer (0.03$\mu$m titanium) is evaporated first, as most seed layer materials (\cite{Seidelin,Labaz}) have a low adhesion on common substrates, see Fig. \ref{CSS} (a). Then a patterned mask (photolithographic structured photo resist) with a negative of the electrode structures is formed on the seed layer. The structures are then formed by electroplating (6$\mu$m gold) metal onto the seed layer, see Fig. \ref{CSS} (b). Afterwards the patterned mask is no longer needed and removed. The seed layer, providing electrical contact between the structures during electroplating, also needs to be removed to allow trapping operations. This can be done using the electroplated electrodes as a mask and an isotropic chemical wet etch process to remove the material, see Fig. \ref{CSS} (c). Then the adhesion layer is removed in a similar etch process (for example using hydrofluoric acid) and the completed trapping structure is shown in Fig. \ref{CSS} (d).\\

\begin{figure}[h]
\begin{center}
\resizebox*{12cm}{!}{\includegraphics{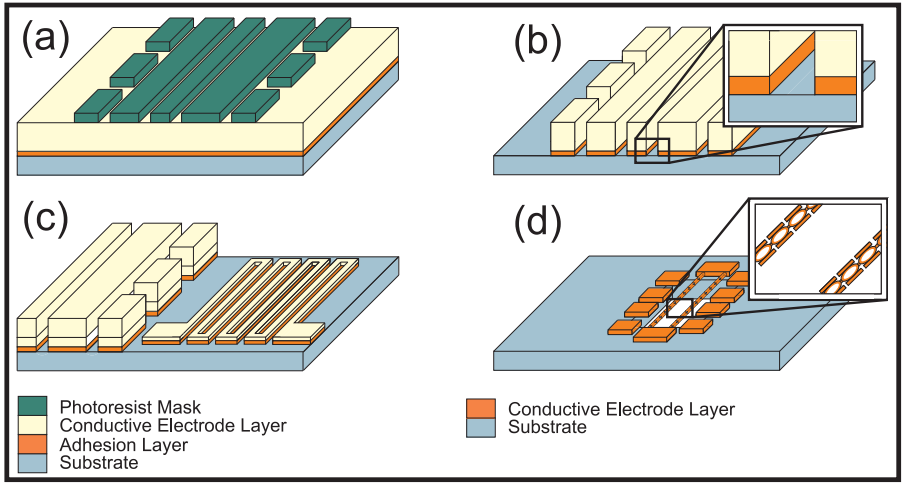}}%
\caption{Different variations of the standard CSS process are shown. (a) Conductive material is deposited on the entire substrate and a resist mask is structured.
(b) Structured electrodes after the etch step. (c) On chip resistor meander line, reported in ref. \cite{Seidelin}.
(d) Variation without exposed dielectrics underneath the trapping zone with free hanging rf rails \cite{StickThesis} .}%
\label{CSS2}
\end{center}
\end{figure}

Depending on the available equipment additional or different steps can be performed, which also allow special features to be included on the chip. One variation was published in \cite{Labaz} replacing the seed layer with a thicker metal layer (1$\mu$m silver used in \cite{Labaz}) and forming the patterned mask on top of that, as shown in Fig. \ref{CSS2} (a). The electrode structures are then formed by using this mask to wet etch through the thick silver layer (NH3OH: H2O2 silver etch) and the adhesion layer (hydrofluoric acid) Fig. \ref{CSS2} (b). To counter potentially sharp electrode edges resulting from the wet etch process an annealing step (720$^{\circ} $C to 760$^{\circ} $C for 1 h) was performed to reflux the material and flatten the sharp edges. An ion trap with superconductive electrodes was fabricated using a similar process \cite{Wang2}. Low temperature superconductors Nb and NbN were grown onto the substrate in a sputtering step. Then the electrodes were defined using a mask and an anisotropic etch step. An annealing step was not performed after this etch step. With this variation of the standard process electroplating equipment and compatible materials are not needed to fabricate an ion trap.\\

A possible addition to the process was presented in \cite{Seidelin} incorporating on-chip meander line resistors on the trap as part of the static voltage electrode filters. Bringing filters closer to the electrodes can reduce the noise induced in connecting wires as the filters are typically located outside the vacuum system or on the chip carrier. On-chip integration of trap features is essential for the future development of very large scale ion trap arrays with controls for thousands of electrodes need for quantum computing.\\

In order to allow for appropriate resistance values for the resistors, processing of the chip occurs in two stages. Within the first stage only the electrode geometry is patterned with regions where resistors are to be fabricated entirely coated in photoresist. This step consists of patterning photo resist on the substrate (for example polished fused quartz \cite{Seidelin}) to shape the electrodes followed by the deposition of the standard adhesion (0.030$\mu$m titanium) and seed layers (0.100$\mu$m copper) on the substrate (thicknesses stated correspond to the values used by Seidelin et al. \cite{Seidelin}). After the electrodes have been patterned, the first mask is removed. In the next step, only the on-chip resistors are patterned allowing for different thickness of conducting layers used for resistors. A second mask is patterned on the substrate parts reserved for the on-chip resistors. Then an adhesion layer (0.013$\mu$m  titanium) and metal layer (0.030$\mu$m gold) is deposited forming the meander lines as shown in Fig. \ref{CSS2} (c). The mask is removed and the rest of the sequence follows the standard process steps. This process illustrates one example for on-chip features integrated on the trap. Another process was published in \cite{Wang}, which integrates on chip magnetic field coils to generate a magnetic field gradient at the trapping zone. It was fabricated using the standard process and a specific mask to form the electrodes incorporating the coils.\\

Most published variations \cite{Seidelin,Wang,Wang2,Labaz,Labaz2,StickThesis,Allcock,Dani} of this process feature structure sizes much smaller than achievable with the PCB process but similar to it electrodes sit directly on the substrate resulting in the same rf loss tangent requirements. It also makes the process incompatible with symmetric ion traps. As a result of the smaller structure separations, high flash over and bulk breakdown voltages are also more important. Because no slots are milled into the substrate electric charges can accumulate on the dielectrics beneath the trapping zones, which can result in stray fields and additional micromotion. To reduce this effect a high aspect ratio of electrode height to electrode-electrode distance is desirable.\\

Similar to the one-layer PCB process, topologies including junctions are possible but isolated electrodes are prohibited by the lack of buried wires. By moving towards cleanroom fabrication techniques it is possible with this process to reduce the electrode - electrode distance and increase the trapping zone density. The scalability is still limited and the exposed dielectrics can lead to unwanted stray fields. To counter the exposed dielectrics another variation and a different process featuring ``patterned silicon on insulator (SOI) layers" can be used and will be explained in Section \ref{SOI}.\\

One variation of this process that prevents exposed dielectrics was fabricated by Sandia National Laboratory \cite{StickThesis} featuring free standing rf rails with a large section of the substrate underneath the trapping region being removed. Electrodes consist of free standing wires held in place by anchor like structures fabricated on the ends of the slot in the substrate as shown in Fig. \ref{CSS2} (d). To prevent snapping under stress the wires are made from connected circles increasing the flexibility. While no dielectrics are exposed underneath the designs scalability and compatibility with different electrode geometries is limited as the rf rails are suspended between the anchors in a straight line.

\subsection{Patterned Silicon on Insulator Layers (SOI)}\label{SOI}

The ``Patterned Silicon on Insulator Layers" process makes use of commercially available silicon-on-insulator (SOI) substrates and was first reported by Britton et al. \cite{Britton2}. Similar to the PCB process this technique removes parts of a substrate instead of adding material to form the ion trap structures. Using the selective etch characteristics of the oxide layer between the two conductive silicon layers of the substrate, it is also possible to create an undercut of the dielectric and shielding the ion from it, without introducing several process steps. A metal deposition step performed at the end of the process to lower the electrode resistance does not require an additional mask, keeping the number of process steps low. Therefore this process allows for much more advanced trap structures without increasing the number of process steps by making clever use of a substrate.\\

\begin{figure}[h]
\begin{center}
\resizebox*{12cm}{!}{\includegraphics{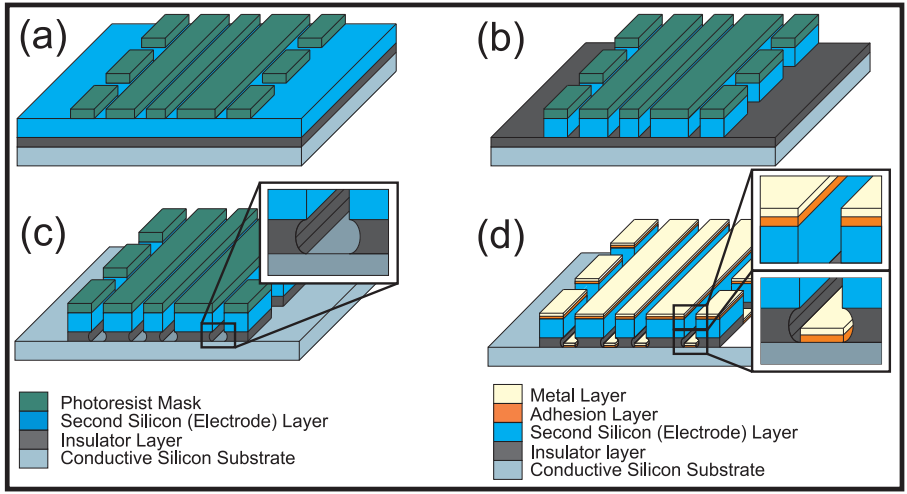}}%
\caption{Fabrication sequence of the SOI process. (a) Resist mask is directly deposited on the substrate. (b) First silicon layer is etched in an anisotropic step.
(c) Isotropic selective wet etch step removing the SiO$_{2}$ layer and creating an undercut. (d) Exposed Si layers coated with adhesion and conductive layer to reduce overall resistance.}%
\label{NIST2009a}
\end{center}
\end{figure}

The substrates are available with insulator thicknesses of up to 10$\mu$m and different Si doping grades resulting in different resistances. After a substrate with desired characteristics (100$\mu$m Si, 3$\mu$m SiO$_{2}$, 540$\mu$m Si in ref. \cite{Britton2}) is found, the process sequence starts with photolithographic patterning of a mask on the top SOI silicon layer as shown in Fig. \ref{NIST2009a} (a). The mask is used to etch through the top Si layer, see Fig. \ref{NIST2009a} (b) by means of an anisotropic process (Deep Reactive Ion Etching (DRIE)). Using a wet etch process the exposed SiO$_{2}$ layer parts are removed and an undercut is formed to further reduce exposed dielectrics as shown in Fig. \ref{NIST2009a} (c). If the doping grade of the top Si layer is high enough the created structure can already be used to trap ions.\\

To reduce the resistance of the trap electrodes resulting from the use of bare Si a deposition step can be used to apply an additional metal coating onto the electrodes. No mask is required for this deposition step, the adhesion layer (Chromium or Titanium) and metal layer (1 $\mu$m Gold, in \cite{Britton2}) can be deposited directly onto the silicon structures as shown in Fig. \ref{NIST2009a} (d). This way the electrodes and the exposed parts of the second Si layer will be coated. This has the benefit that possible oxidization of the lower silicon layer is also avoided. A slot can be fabricated into the substrate underneath the trapping zone, which allows backside loading in these traps \cite{Britton2}. This slot is microfabricated by means of anisotropic etching using a patterned mask from the backside. The benefit of such a slot is that the atom flux needed to ionize atoms in the trapping region can be created at the backside and guided perpendicular to the trap surface away from the electrodes. Therefore coating of the electrodes as a result of the atom flux can be minimized.\\

An ion trap based on the SOI process was fabricated with shielded dielectrics by Britton et al. \cite{Britton2}. A similar approach is used by Sterling et al. \cite{Sterling} to create 2-D ion trap arrays. Similar to the standard CSS process, discussed in the previous section, y-junctions and other complex topologies are possible, but buried wires are prohibited. The process design incorporates a limited material choice for substrate and insulator layer.  Only the doping grade of the upper and lower conductive layers can be varied not the material. The insulator layer separating the Si layers must be compatible with the SOI fabrication technique and commercially available materials are SiO$_{2}$ and Al$_{2}$O$_{3}$ (Sapphire). Adjustments of electrical characteristics, like rf loss tangent and electrode capacitances are therefore limited to the two insulator materials and geometrical variations.\\

To further increase the scalability and trapping zone density of ion traps, process techniques should incorporate buried wires to allow isolated static voltage and rf electrodes. Examples of such monolithic processes will be discussed next, first a process incorporating buried wires but exhibiting exposed dielectrics will be presented followed by a process allowing for buried wires and shielding of dielectrics.

\subsection{Conductive Structures on Insulator with Buried Wires (CSW)}

\begin{figure}[h]
\begin{center}
\resizebox*{12cm}{!}{\includegraphics{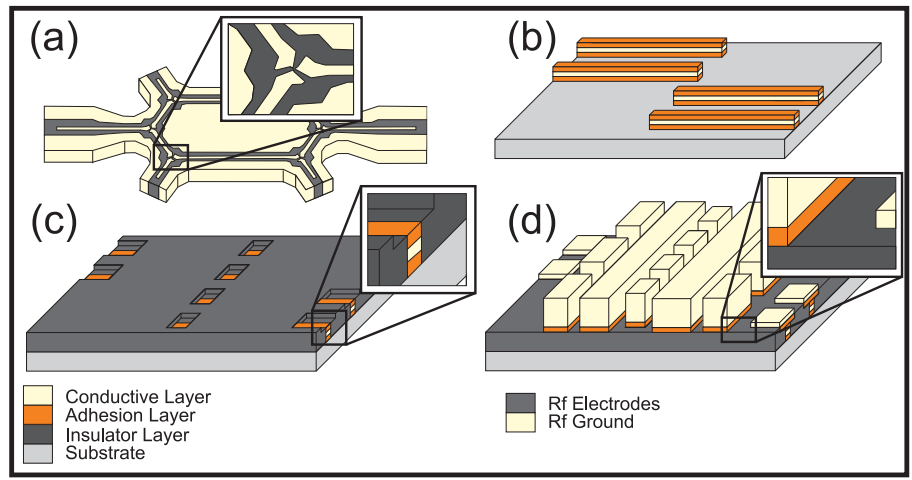}}%
\caption{Process sequence for the buried wire process. (a) Hexagonal shaped six junction ion trap array \cite{Amini}. (b) A structured resist mask is deposited and the mask is used to form wires in a following wet etch step. (c) The wires are then buried with an oxide layer and windows are formed in the oxide. (d) With another mask the electrodes are deposited on top of the wires and oxide layer.}%
\label{NIST2010b}
\end{center}
\end{figure}

The process to be discussed here is a further development of the CSS process discussed in section \ref{CSSPro}, which adds two more layers to incorporate buried wires. With these it is possible to connect isolated static voltage and rf electrodes and was used to fabricate an ion trap with six junctions arranged in a hexagonal shape \cite{Amini}, see Fig. \ref{NIST2010b} (a). The capability of buried wires is essential for more complex ion trapping arrays that could be used to trap hundreds or thousands of ions necessary for advanced quantum computing.\\

The process starts with the deposition of the buried conductor layer (0.300$\mu$m gold layer \cite{Amini}) sandwiched between two adhesion layers (0.02$\mu$m titanium) on the substrate (380$\mu$m quartz) and a patterned mask is deposited on these layers. Conductor and adhesion layers are then patterned in an isotropic etch step, see Fig. \ref{NIST2010b} (b) and then buried with an insulator material (1$\mu$m SiO$_{2}$ deposited by means of chemical vapor deposition (CVD)). To establish contact between electrodes and the buried conductor, windows are formed in the insulator layer (plasma etching) as shown in Fig. \ref{NIST2010b} (c). Then another patterned mask is used to deposit an adhesion and conductor layer forming the electrodes (0.020$\mu$m titanium and 1$\mu$m gold). In this deposition step the windows in the insulator are also filled and connection between buried conductor and electrodes is established as shown in Fig. \ref{NIST2010b} (d). For this ion trap  a backside loading slot was also fabricated using a combination of mechanical drilling and focused ion beam milling to achieve a precise slot in the substrate \cite{Amini}. As demonstrated by Amini et al. \cite{Amini} this process allows for large scale ion trap arrays with many electrodes and trapping zones. While the scalability is further increased, this particular process design results in exposed dielectrics. The next process improves this further by shielding the electrodes and the substrate.

\subsection{Conductive Structures on Insulator with Ground Layer (CSL)}

To achieve buried wires or in this case vertical interconnect access (vias) and shielding of dielectrics a monolithic process including a ground layer and overhanging electrodes can be carried out. Several structured and deposited layers are necessary for this process and while providing the greatest flexibility of all processes it also results in a complicated process sequence. Therefore these processes are commonly performed using Very-Large-Scale Integration (VLSI) facilities that are capable of performing many process steps with high reliability.\\

\begin{figure}[h]
\begin{center}
\resizebox*{12cm}{!}{\includegraphics{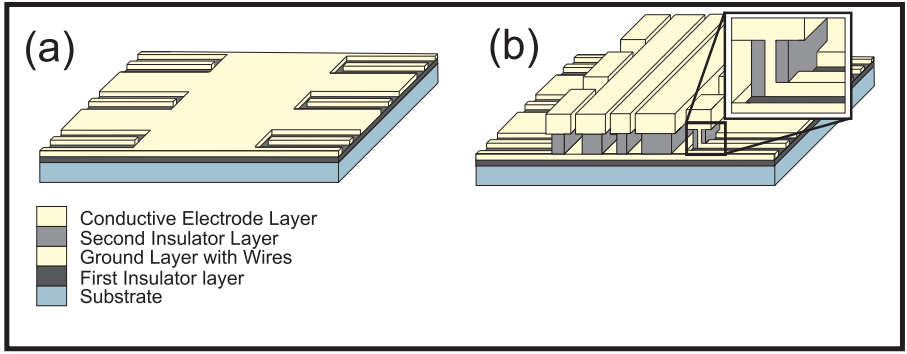}}%
\caption{Two variations of the CSL process sequence. (a) In the process sequence published in \cite{Sandia} an insulating layer and a structured metal ground layer are deposited on the substrate. (b) Electrodes fabricated in one plane, with outer static voltage electrodes connected through via's \cite{Sandia}.}%
\label{Lucentb}
\end{center}
\end{figure}

A process design, which makes use of vias, was presented by Stick et al. \cite{Sandia}. In this process design the substrate (SOI in ref. \cite{Sandia}) is coated with an insulator and a structured conductive ground layer (1$\mu$m Al) as shown in Fig. \ref{Lucentb} (a). This is followed by a thick structured insulator (9-14 $\mu$m). The trapping electrodes are then placed in a plane above the thick insulator. The electrodes overhang the thick insulator layer and in combination with the conductive ground layer shield the trapping zone from dielectrics. Vias are used to connect static electrodes to the conductive layer beneath the thick insulator as shown in Fig. \ref{Lucentb} (b). The process also includes creation of a backside loading slot.\\

A variation from the discussed process making use of a ground plate without vias is described by Leibrandt et al. \cite{Leibrandt} and more detailed process steps are given in ref. \cite{StickThesis}.\\

The described process design \cite{Sandia} combines vias with shielded dielectrics and also makes the trap structures independent from the substrate. Therefore the same level of scalability as the CSW process, discussed in the previous section, can be achieved while dielectrics are shielded similar to the SOI process, described in section \ref{SOI}. The process uses aluminium to form the electrodes which can lead to oxidisation and unwanted charge build up. To avoid this the electrodes could be coated with gold or other non-oxidizing conductors in an additional step. Easier choice of substrate, isolator and electrode materials combined with vias and shielded dielectrics make this process well suited for very large scale asymmetric ion traps with high trapping zone densities and low stray fields. With electrodes in one plane and another conductive layer above the substrate this structure could also be used to fabricate a symmetric ion traps (SIT).

\subsection{Double Conductor/Insulator Structures on Substrate (DCI)}

Symmetric ion traps have electrodes placed in two or three planes with the ions trapped between the planes. Therefore electrodes need to be precisely structured in several vertically separated layers and cannot be fabricated in one plane resulting in different requirements on the microfabrication processes. In the DCI process described here a specifically grown substrate with selective etch capabilities similar to SOI substrate layers is used and all electrodes are structured in one etch steps. The first such microfabricated ion trap was created by Stick et al. \cite{Stick} using this process with MBE grown AlGaAs, GaAs structures and constitutes the first realisation of a monolithic ion trap chip.\\

\begin{figure}[h]
\begin{center}
\resizebox*{12cm}{!}{\includegraphics{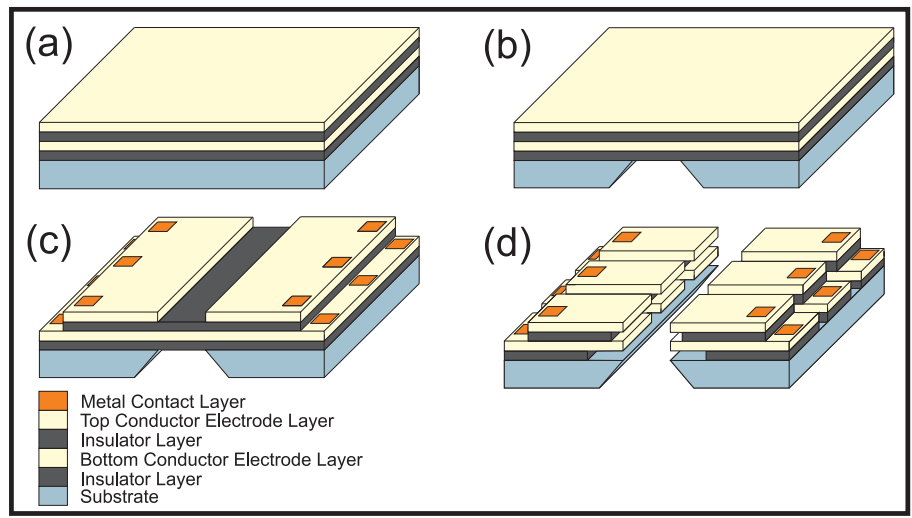}}%
\caption{Fabrication process used for the symmetric trap published in \cite{Stick}. (a) Alternately grown conductive and insulating layers on a substrate.
(b) Backside wet etch step to allow for optical access of the trapping zone. (c) Structured top GaAs and AlGaAs layers, formed in an etch step using a mask. Metal contacts deposited on both electrode layers.
(d) Structured top and bottom electrodes separated by the insulator layers and undercuts are formed by a selective wet etch. }%
\label{DCI}
\end{center}
\end{figure}

The process starts by growing alternating layers (AlGaAs 4$\mu$m, GaAs 2.3$\mu$m) on a substrate using an MBE system  as shown in Fig. \ref{DCI} (a). The doping grades and atomic percentages (70 \% Al, 30\% GaAs AlGaAs layer, GaAs layers highly doped $3*10^{18} /cm^{3}$) are chosen to achieve low power dissipation in the electrodes. After the structure is grown a slot is etched into the substrate from the backside to allow optical access as shown in Fig. \ref{DCI} (b). To provide electrical contact to both GaAs layers an anisotropic etch process is performed to remove parts of the first AlGaAs and GaAs layers, then metal is deposited onto parts of the top GaAs and the now exposed lower GaAs layer as shown in Fig. \ref{DCI} (c). This is followed by an anisotropic etch step defining all the electrode structures. A following isotropic etch step creates an undercut of the insulating AlGaAs layer to increase the distance between trapping zone and dielectric, see Fig. \ref{DCI} (d).\\

This ion trap chip could only be operated at very low rf amplitudes of 8V as a result of the low breakdown voltage of the insulating AlGaAs material and possible residues on the insulator surfaces. This problem can be addressed by choosing a different and/or thicker insulator material.\\

Removing the insulator material between the electrode layers can avoid this problem almost entirely and was used in the design described by Hensinger et al. \cite{Winni}. This structure is based on depositing layers instead of using a specifically grown substrate and etching material away. It incorporates a selectively etchable oxide layer, which can be completely removed after electrodes are created on this layer. This sacrifical layer allows for cantilever like electrodes without any supporting oxide layers. Therefore breakdown can only occur over the surfaces and the capacitance between electrodes is kept at a minimum.\\

\begin{figure}[h]
\begin{center}
\resizebox*{12cm}{!}{\includegraphics{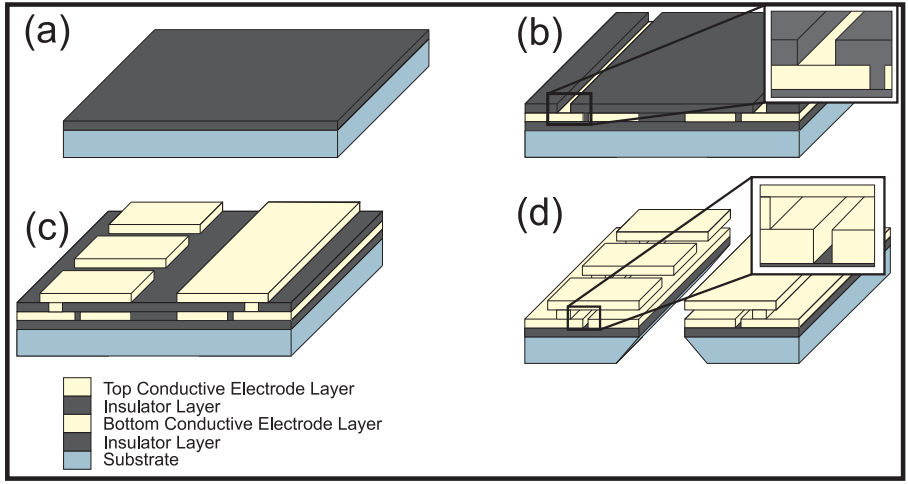}}%
\caption{Process sequence for the trap proposed in \cite{Winni}. (a) Insulator layer deposited on the substrate. (b) Structured electrode conductor layer and a second selectively etchable insulator is deposited and structured. (c) The second conductor layer is deposited, making a vertical electrical connection to the lower conductor layer and is structured using an anisotropic etch process. (d) All selectively etchable insulator layers are removed and a backside etch is performed creating a slot in the substrate.}%
\label{DCI2}
\end{center}
\end{figure}

The process starts with the deposition of an insulating layer (Si$_{3}$N$_{4}$) on an oxidized Si substrate as shown in Fig. \ref{DCI2} (a). Then a conductor layer (polycrystal silicon \cite{Winni}) is deposited onto the insulator and structured using an anisotropic etch process (DRIE) with a mask. Now selectively etchable sacrificial material (polysilicon glass) is deposited on the conductor structures. With a second mask windows are etched into the insulator layer allowing vertical connections to the following upper electrodes, see Fig. \ref{DCI2} (b). This is followed by another conductor layer (polycrystalline silicon) deposited onto the patterned insulator also filling the previously etched windows. Then an anisotropic etch is performed to structure the top conductor layer as shown in Fig. \ref{DCI2} (c). An etch step using a mask is then performed from the backside through the entire substrate creating a slot in the silicon substrate. With an isotropic selective hydrofluoric etch the sacrificial material used to support the polycrystal silicon electrodes during fabrication is completely removed. The electrodes now form cantilever like structures and the previously created slot in the substrate allows optical access to the trapping zone, see Fig. \ref{DCI2} (d).\\

This design potentially allows for the application of much higher voltages (breakdown for this design would occur over an insulator surface rather than through insulator bulk).\\

Independent of the material, the distance between the two electrode planes is limited by the maximal allowable insulator thickness which is often determined by particular deposition and etch constraints. To keep the ion-electrode distance at a desired level the aspect ratio of horizontal and vertical electrode-electrode distances (see Fig. \ref{DCI2} (d)) has to be much higher than one. As described in chapter \ref{linear} this leads to a low geometric efficiency factor $\eta$ and reduces the trap depths and secular frequencies of these traps. By forming the electrodes on the upper and lower side of an oxidized substrate this aspect ratio can be dramatically decreased. A process technique following this consideration will be discussed next.

\subsection{Double Conductor Structures on Oxidized Silicon Substrate (DCS)}

\begin{figure}[h]
\begin{center}
\resizebox*{12cm}{!}{\includegraphics{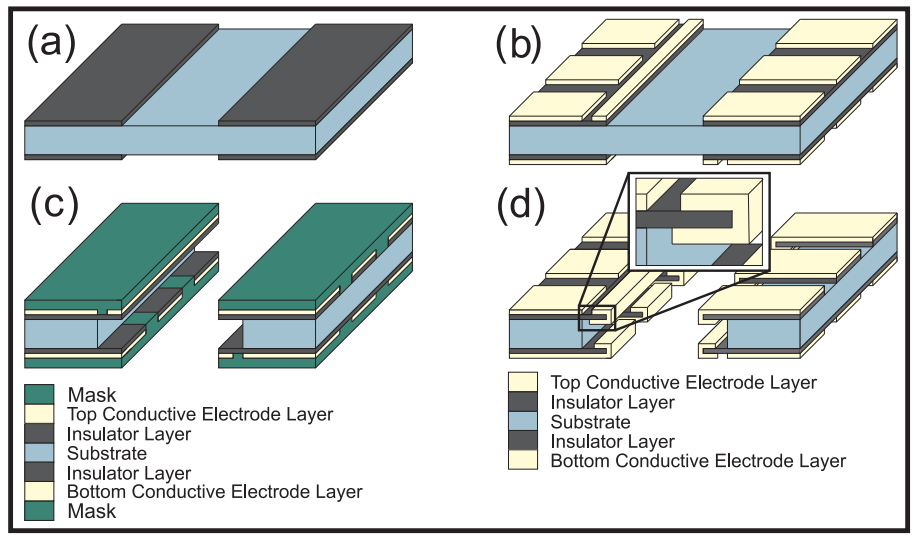}}%
\caption{Process sequence as used by Brownnutt et al. \cite{Brownnutt}. (a) A structured insulator layer is deposited on both sides of the substrate. (b) Electrode structures are then deposited on this insulator layer.
(c) Parts of the insulator are also removed in this step and a slot is etched into the substrate. The top and bottom structures are then covered with a mask.
(d) Shadow evaporation is used to coat exposed dielectrics close to the trapping zone. The mask is removed and the conductor layer thickness increased by means of electroplating. }%
\label{DCS}
\end{center}
\end{figure}

In this proposed monolithic process \cite{Brownnutt} an oxidized silicon substrate separates the electrode structures, which allows for much higher electrode to electrode distances. Making use of both sides of the substrate separates this process from all other described techniques. It allows electrode-electrode distances of hundreds of micrometers, which would otherwise be impossible due to maximal thicknesses of deposited or grown oxide layers. Therefore a ratio of one for the horizontal and vertical electrode-electrode distances can be achieved, resulting in an optimal geometric efficiency factor $\eta$. The trap is fabricated in several process steps starting with the oxidization of a silicon wafer. Using an anisotropic etch process and a patterned mask, slots are then formed in the SiO$_{2}$ layers on both sides exposing the Si wafer and creating the future trenches between electrodes as shown in Fig. \ref{DCS} (a). Then conductive layers are deposited on both sides of the wafer forming the two electrode layers. The individual electrodes are then created on both sides in an anisotropic etch step using a patterned mask as shown in Fig. \ref{DCS} (b). This is followed by an isotropic wet etch through the silicon using another mask resulting in an under cut of the electrodes and providing the optical access as shown in Fig. \ref{DCS} (c). The now exposed dielectric SiO$_{2}$ layers are then coated in a shadow metal evaporation process and the remaining mask are removed. After all conductor layers are deposited, the layer thickness is increased by means of electroplating  resulting in the final structure shown in Fig. \ref{DCS} (d) \cite{Brownnutt}. By using the entire substrate to separate the electrodes this proposed process sequence allows for a low aspect ratio of horizontal and vertical electrode electrode distance. All electrodes sit on a dielectric material increasing the importance of a low rf loss tangent of the insulating SiO$_{2}$ layers.

\subsection{Assembly of Precision Machined structures (PMS)}

Non-monolithic fabrication processes are commonly based on the assembly of pre-structured parts, using techniques like wafer bonding or mechanical clamping. The scalability of non-monolithic processes can be limited due to this and therefore are commonly used for ion traps with only one or a few trapping zones. Systems with thousands of isolated electrodes would be difficult to realize with a non-monolithic process. Advantages of the non-monolithic processes are a greater choice of materials used and fabrication techniques, ranging from machined metal structures over laser machined alumina, to structured Si substrates.\\

A non-monolithic process can be based on the assembly of precision machined structures commonly done in workshops and will not be discussed in detail. Two examples for ion traps fabricated with this process are the needle trap with needle tip radius of approximately 3$\mu$m reported by Deslauriers et al. \cite{D06} and a blade trap used by McLoughlin et al. in \cite{McLoughlin}. More complex ion trap geometries including junctions and more electrodes can also be realized with a non-monolithic processes and will therefore be discussed next.

\subsection{Assembly of Laser Machined Alumina structures (LMA)}

\begin{figure}[h]
\begin{center}
\resizebox*{12cm}{!}{\includegraphics{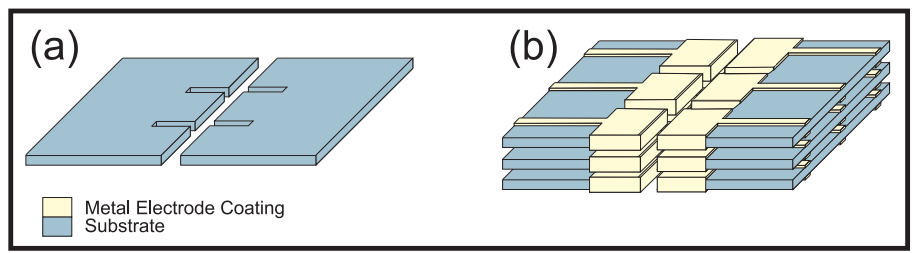}}%
\caption{``Assembly of Laser Machined Alumina structures" process sequence. (a) Cantilever like structures are formed into the alumina structures using a laser. (b) Parts of the substrate are coated using a mask forming the electrodes and electrical connections. Three similar layers are mounted on top of each other to form a three layer symmetric trap.}%
\label{LMA}
\end{center}
\end{figure}

Trap structures can be created by precision laser machining and coating of alumina substrates and mechanically assembling these using spacers and clamps as shown in Fig. \ref{LMA}. The structures are created by laser machining slots into an alumina substrate resulting in cantilevers providing the mechanical stability for the electrodes, see Fig. \ref{LMA} (a). The alumina cantilevers are then coated with a conductive layer, commonly metal, to form the electrodes. To prevent electrical shorting between the electrodes a patterned mask is used during this step. Several structured and coated alumina plates are then mechanically assembled to form a two or three layer trap \cite{T00,Hensinger,Blakestad}. Normally Spacers are used to maintain an exact distance between the plates, which are then exactly positioned and held in place using mechanical pressure, see Fig. \ref{LMA} (b).\\

One of the first ion traps with linear sections using this process was reported by Turchette et al. \cite{T00}. Another example for such a trap was used for the first demonstration of corner shuttling of ions within a two-dimensional ion trap array \cite{Hensinger}. Other traps fabricated with this process include the trap used for near adiabatic shuttling through a junction \cite{Blakestad} and the traps reported in refs. \cite{Rowe,Schulz}. The necessary alumina substrates show a small loss tangent. A similar non-monolithic process based on clean room fabrication techniques, which allows for higher precision and greater choice of substrate material, will be discussed next.

\subsection{Wafer Bonding of Lithographic Structured Semiconductor Substrates (WBS)}
This process technique is similar to the monolithic SOI substrate technique discussed in section \ref{SOI} creating a highly doped conductor on an insulator using wafer bonding techniques. Much higher insulator thicknesses can be achieved with this process and many types of insulator and substrates can be used. This process was used to fabricate symmetric and asymmetric ion traps.\\
\begin{figure}[h]
\begin{center}
\resizebox*{12cm}{!}{\includegraphics{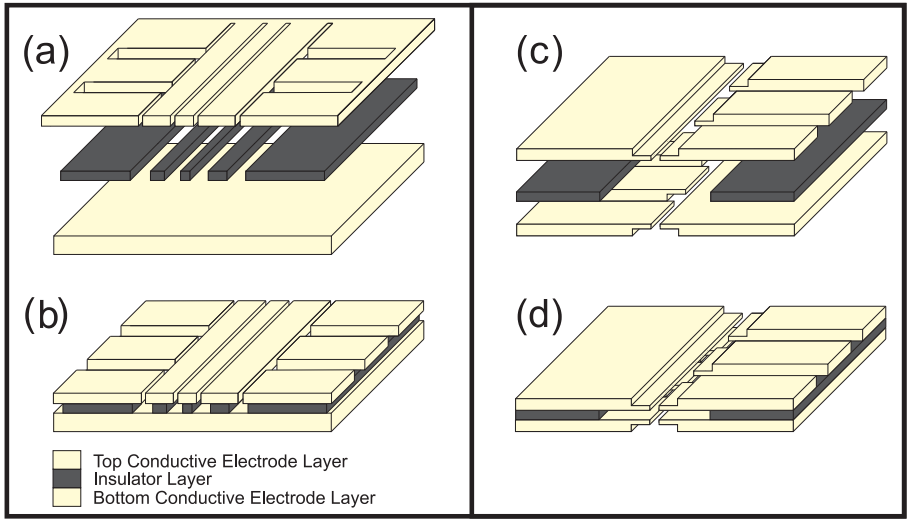}}%
\caption{Processes used to fabricate symmetric and asymmetric non-monolithic ion traps via waver bonding. (a) Microfabricated structures used for the assembly of an asymmetric trap. The structures are still physically connected within each layer.
(b) These parts are then waferbonded together and physical connections are removed forming the trap structure. (c) Microfabricated parts used for a symmetric trap. (d) Waferbonded parts creating a symmetric trap. }%
\label{WBS}
\end{center}
\end{figure}

First a process used for an asymmetric ion trap will be discussed, starting with a commercially available substrate (Si wafer, resistivity$ = 500 \times 10^{-6} \Omega$ cm). Using anisotropic etching with a patterned mask (DRIE) the wafer is structured. All electrodes are physically connected outside the trapping zone to provide the necessary stability during the fabrication as shown in Fig. \ref{WBS} (a). The structured substrate is then bonded on another substrate with a sandwiched and structured insulator layer providing electrical insulation and mechanical stability as shown in Fig. \ref{WBS} (b). In the insulator layer a gap is formed underneath the trapping region leaving no exposed dielectrics underneath the ion. The bonding also provides the needed structural stability and the physical connections between the electrodes can then be removed by dicing the substrate Fig. \ref{WBS} (b). Similar to the SOI substrate process a conductive layer can be deposited on top of the structured substrate. A patterned mask is not necessary and the conductive layer can be directly deposited. Depending on the material used and substrate an adhesion layer has to be added.\\

Another variation of this process can be used to fabricate symmetric ion traps. Both substrates are identically structured and wafer bonded. An additional etch step is introduced to make the substrate thinner adjacent to the trapping zone in order to improve optical access as shown in Fig. \ref{WBS} (c). Similar to the asymmetric trap, the parts are then wafer bonded together with a sandwiched insulator layer Fig. \ref{WBS} (d). When used for asymmetric traps the dielectrics are completely shielded from the trapping zone and for the case of symmetric traps the dielectrics can be placed far away from the trapping zone. Possible geometries for asymmetric ion traps are limited with this technique as buried wires and therefore isolated electrodes are not possible. The fabrication method and choice of electrode materials used for this trap can result in a very low surface roughness of less than 1nm.\\

\section{Anomalous heating}\label{heating}
One limiting factor in producing smaller and smaller ion traps is motional heating of trapped ions. While it only has limited impact in larger ion traps it becomes more important when scaling down to very small ion - electrode distances. The most basic constraint is to allow for laser cooling of the ion. If the motional heating rate is of similar magnitude as the photon scattering rate, laser cooling is no longer possible. Therefore for most applications, the ion - electrode distance should be chosen so the expected motional heating rate is well below the photon scattering rate. Depending on the particular application, there may be more stringent constraints. For example, in order to realize high fidelity quantum gates that rely on motional excitation for entanglement creation of internal states \cite{molmer,PLee}, motional heating should be negligible on the time scale of the quantum gate. This timescale is typically related to the secular period $1/\omega_m$ of the ion motion, however, can also be faster \cite{Garcia}.  Motional heating of trapped ions in an ion trap is caused by fluctuating electric fields (typically at the secular frequency of the ion motion). These electric fields originate from voltage fluctuations on the ion trap electrodes. One would expect some voltage fluctuations from the electrodes due to the finite impedance of the trap electrodes, this effect is known as Johnson noise. Resulting heating would have a $1/d^2$ scaling \cite{T00} where $d$ is the characteristic nearest ion-electrode distance. However, in actual experiments a much larger heating rate has been observed. In fact, heating measurements taken for a variety of ions and ion trap materials seem to loosely imply a $1/d^4$ dependence of the motional heating rate $\dot{\bar{n}}$. A mechanism beyond Johnson noise must be responsible for this heating and this mechanism was termed 'anomalous heating'. In order to establish a more reliable scaling law, an experiment was carried out where the heating rate of an ion trapped between two needle electrodes was measured \cite{D06}. The experimental setup allowed for controlled movement of the needle electrodes. It was therefore possible to vary the ion - electrode distance and an experimental scaling law was measured $\dot{\bar{n}}\sim1/d^{3.5\pm0.1}$ \cite{D06}. The motional heating of the secular motion of the ion can be expressed as \cite{T00}
\begin{equation}\label{ndot}
\dot{\bar{n}}=\frac{q^2}{4m\hbar \omega_m}\left(S_E(\omega_m)+\frac{\omega^2_m}{2\Omega^2_T}S_E(\Omega_T \pm \omega_m)\right)
\end{equation}
$\omega_m$ is the secular frequency of the mode of interest, typically along the axial direction of the trap, $\Omega_T$ is the drive frequency and the power spectrum of the electric field noise is defined as $S_E(\omega)=\int^\infty_{-\infty}\langle E(\tau)E(t+\tau)\rangle e^{i\omega\tau}d\tau$. The second term represents the cross coupling between the noise and rf fields and can be neglected for axial motion in linear traps as the axial confinement is only produced via static fields \cite{Wineland,T00}.\\

\begin{figure}[htp]
\begin{center}
\resizebox*{15cm}{!}{\includegraphics{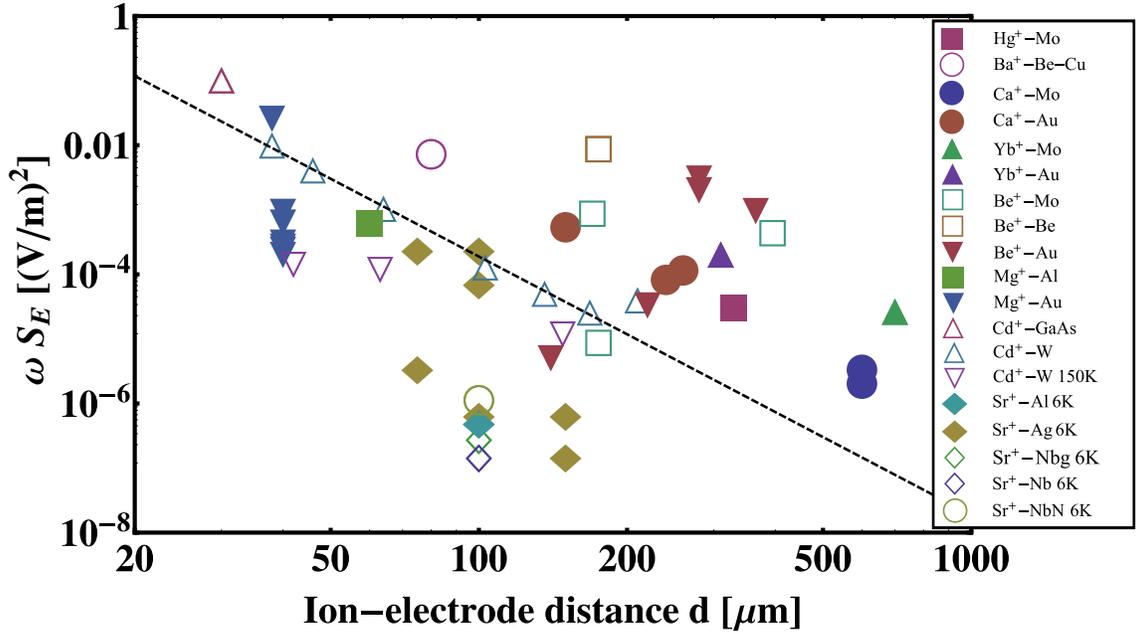}}%
\caption{Previously published measurements of motional heating plotted as the product of electric field noise spectral density $S_{E}(\omega)$ and the secular frequency $\omega$, versus ion-electrode distance d.
A $1/d^4$ trend line is also shown. Each label shows both the ion species and the electrode material used and the electrode temperature is also noted if the measurement is performed below room temperature. The data point are associated with the following references.
(Hg$^{+}-$Mo \cite{Diedrich}, Ba$^{+}-$Be-Cu \cite{DeVoe}, Ca$^{+}-$Mo \cite{Roos}, Ca$^{+}-$Au \cite{Schulz,Allcock,Dani2}, Yb$^{+}-$Mo \cite{Tamm}, Yb$^{+}-$Au \cite{McLoughlin}, Be$^{+}-$Mo \cite{T00}, Be$^{+}-$Be \cite{T00}, Be$^{+}-$Au \cite{T00,Rowe}, Mg$^{+}-$Al \cite{BrittonThesis}, Mg$^{+}-$Au \cite{Epstein,Seidelin,Amini,Britton2,BrittonThesis}, Cd$^{+}-$GaAs \cite{Stick}, Cd$^{+}-$W \cite{D06}, Cd$^{+}-$W 150K \cite{D06}, Sr$^{+}-$Al 6K \cite{Wang2}, Sr$^{+}-$Ag 6K \cite{Labaz}, Sr$^{+}-$Nbg 6K, Sr$^{+}-$Nb 6K, Sr$^{+}-$NbN 6K \cite{Wang2})}
\label{EFN}
\end{center}
\end{figure}

A model was suggested to explain the $1/d^4$ trend that considered fluctuating patch potentials; a large number of randomly distributed `small' patches on the inside of a sphere, where the ion sits at centre at a distance $d$ \cite{T00}. All patches have a power noise spectral density that influence the electric field at the ion position, over which is averaged to eventually deduce the heating rate. Figure \ref{EFN} shows a collection of published motional heating results. Instead of plotting the actual heating rate, we plot the spectral noise density $S_E(\omega_m)$ multiplied with the secular frequency in order to scale out behavior from different ion mass or different secular frequencies used in individual experiments. We also plot a $1/d^4$ trend line. We note that previous experiments \cite{D06,McLoughlin} consistently showed  $S_E(\omega_m)\sim 1/\omega_m$ allowing the secular frequency to be scaled out by plotting $S_E(\omega_m)\times\omega_m$ rather than just $S_E(\omega_m)$.
In the experiment by Deslauriers et al. \cite{D06}, another discovery was made. The heating rate was found to be massively suppressed by mild cooling of the trap electrodes. Cooling the ion trap from 300K down to 150K reduced the heating by an order of magnitude \cite{D06}. This suggests the patches are thermally activated. Labaziewicz et al. \cite{Labaz} measured motional heating for temperatures as low as 7K and found a multiple-order-of-magnitude reduction of motional heating at low temperatures. The same group measured a scaling law for the temperature dependence of the spectral noise density for a particular ion trap as $S_E(T)= 42(1+(T/46 K)^{4.1})\times10^{-15}$ V$^2$/m$^2$/Hz \cite{Labaz2}.
Superconducting ion traps consisting of niobium and niobium nitride were tested above and below the critical temperature $T_c$ and showed no significant change in heating rate between the two states \cite{Wang2}. Within the same study, heating rates were reported for gold and silver trap electrodes at the same temperature (6 K) showing no significant difference between the two and superconducting electrodes. This suggests that anomalous heating is mainly caused by noise sources on the surfaces, although the exact cause of anomalous heating is not yet fully understood. From the information available, it is likely that surface properties play a critical role, however, other factors such as material bulk properties, oxide layers may also play an important role and much more work is still needed to fully understand and control anomalous heating. While anomalous heating limits our ability to make extremely small ion traps, it does not prevent the use of slightly larger microfabricated ion traps. Learning how to mitigate anomalous heating is therefore not a prerequisite for many experiments, however, mitigating it will help to increase experimental fidelities (such as in quantum gates) and will allow for the use of smaller ion traps.

\section{Conclusion}
Microfabricated ion traps provide the opportunity for significant advances in quantum information processing, quantum simulation, cavity QED, quantum hybrid systems, precision measurements and many other areas of modern physics. We have discussed the basic principles of microfabricated ion traps and highlighted important factors when designing such ion traps. We have discussed important electrical and material considerations when scaling down trap dimensions. We presented a detailed overview of a vast range of ion trap geometries and showed how they can be fabricated using different fabrication processes employing advanced microfabrication methods. A limiting factor to scaling down trap dimensions even further is motional heating in ion traps and we have summarized current knowledge of its nature and how it can be mitigated. The investigation of microfabricated ion trap lies on the interface of atomic physics and state-of-the-art nanoscience. This is a very young research field with room for many step-changing innovations. In addition to the development of trap structures themselves, future research will focus on advanced on-chip features such as cavities, electronics, digital signal processing, fibres, waveguides and other integrated functionalities.  Eventually progress in this field will result in on-chip architectures for next generation quantum technologies allowing for the implementation of large scale quantum simulations and quantum algorithms. The inherent scalability of such condensed matter systems coupled with the provision of atomic qubits within which are highly decoupled from the environment will allow for ground-breaking innovations in many areas of modern physics.

\section*{Acknowledgements}
We acknowledge the support by the UK Engineering and Physical Sciences Research Council (EP/E011136/1, EP/G007276/1), the European Commission's Sixth Framework Marie Curie International Reintegration Programme (MIRG-CT-2007-046432), the Nuffield Foundation and the University of Sussex.

\newpage

\section*{Notes on contributors}

\begin{figure}[!h]
\begin{center}
\resizebox*{3cm}{!}{\includegraphics{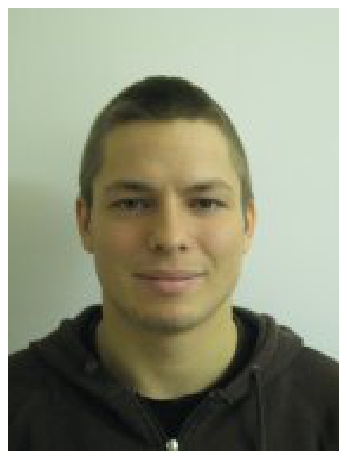}}
\\{Marcus Hughes gained his MPhys Physics at the University of Sussex in 2008 and has carried on studying for his D.Phil. within the Sussex Ion Quantum Technology group. His interests include quantum information and microfabricated ion traps.\\}
\end{center}
\end{figure}

\begin{figure}[!h]
\begin{center}
\resizebox*{3cm}{!}{\includegraphics{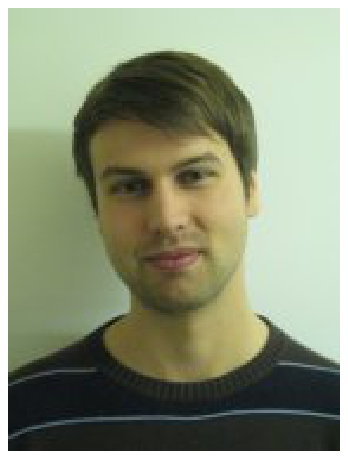}}%
\\{Bjoern Lekitsch is a D.Phil student at the University of Sussex in the Ion Quantum Technology group. He completed his Dipl. Ing. degree in Nanostrukturtechnik at the Physics Department at the University of Wuerzburg, Germany in 2010. His interests include quantum information technology, microfabrication of ion traps and shuttling of ions trough optimized junctions in large ion traps arrays.\\}
\end{center}
\end{figure}

\begin{figure}[!h]
\begin{center}
\resizebox*{3cm}{!}{\includegraphics{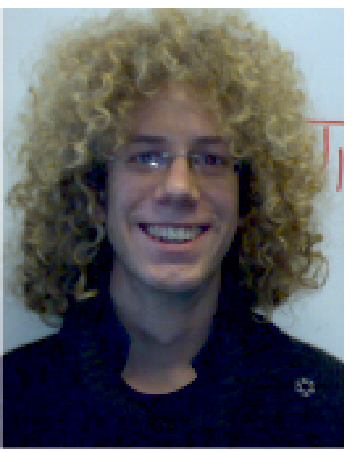}}%
\\{Jiddu Broersma, a University of Sussex graduate student obtained his MPhys Theoretical Physics in 2010. He completed his final year research project in the Sussex Ion Quantum Technology group. His interests include quantum information science, neural networks and renewable energies.\\}
\end{center}
\end{figure}

\newpage

\begin{figure}[!h]
\begin{center}
\resizebox*{3cm}{!}{\includegraphics{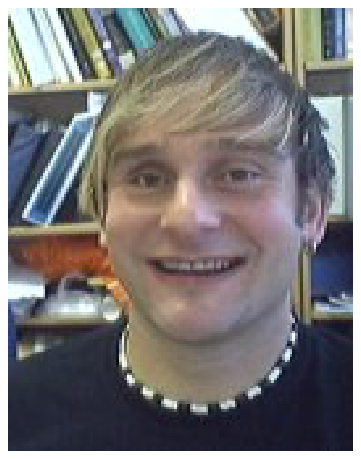}}%
\\{Winfried Hensinger is a Reader in Quantum, Atomic and Optical Physics and heads the Ion Quantum Technology group at the University of Sussex. In 2008 he was awarded an EPSRC Leadership fellowship for the development of quantum technology with nanofabricated ion trap chips. He obtained his undergraduate degree at the Ruprechts-Karls University in Heidelberg, Germany and his PhD at the University of Queensland in Brisbane, Australia studying experimental quantum nonlinear dynamics with ultracold atoms. During his PhD candidature he spent an extended period at the National Institute of Standards and Technology in Gaithersburg, USA where he demonstrated dynamical tunnelling in a Bose-Einstein condensate. After completing his PhD he spent three years as a FOCUS Research Fellow at the University of Michigan, USA working on scalable quantum information processing with trapped ions. He developed the first monolithic ion trap chip and demonstrated the first experimental realization of a two-dimensional ion trap array where individual ions can be shuttled around corners.  Hensinger's primary research direction is the development of quantum technologies beyond proof-of-principle.\\}
\end{center}
\end{figure}

\end{document}